\documentclass[aps,physrev,preprint,groupedaddress]{revtex4-2}

\usepackage{graphicx,amsmath, amssymb, amsthm,subcaption,hyperref,braket,dsfont,mathrsfs}
\begin{document}


\title{Geometric-Symmetry Logical Gate and Local-Probe Selectivity in the Three-Leg AKLT Ladder}


\author{Jingnuo Han}
\affiliation{Department of Applied Physics, College of Science,
China Agricultural University, Qinghua East Road, Beijing 100083, the People’s Republic of China}

\author{Youning Li}
\email[]{liyn03@gmail.com}
\affiliation{Department of Applied Physics, College of Science,
China Agricultural University, Qinghua East Road, Beijing 100083, the People’s Republic of China}


\date{\today}

\begin{abstract}
Symmetry-protected topological (SPT) phases provide a platform for encoding quantum information in protected boundary degrees of freedom.
Here we study the three-leg Affleck-Kennedy-Lieb-Tasaki (AKLT) ladder as an exactly solvable SPT system with on-site symmetry $SO(3)\times \mathbb{Z}_2$.
Using an exact matrix product state construction, we characterize the symmetry action on the edge encoding space and the accessibility of this space by local operators.
We find that the continuous $SO(3)$ symmetry induces boundary rotations, while the leg-exchange symmetry generates a geometry-dependent logical permutation of edge qubits.
Furthermore, by introducing a distinguishability measure motivated by the Knill--Laflamme condition, we derive a symmetry-resolved decay law for local accessibility.
The decay is controlled by a selection rule raised from the Wigner--Eckart
theorem, whereby a rank-$\ell$ local operator couples only to the $\mathcal L=\ell$ transfer-matrix sector, with a decay length determined by the corresponding correlation length.
We further identify a finite-size channel that is independent of the probe operator position.
These results establish a quantitative connection between SPT symmetry, lattice geometry, and the protection of boundary-encoded quantum information.
\end{abstract}
\maketitle
\section{Introduction}
\label{sec:intro}

The potential of quantum computing lies in its ability to exploit quantum superposition and entanglement to solve problems that are intractable for classical computers. However, this potential is hindered by a fundamental obstacle: quantum states are extremely fragile. Therefore, reliably encoding and manipulating quantum information in physical systems remains one of the central challenges in quantum computing.
The standard approach is active quantum error correction, in which syndrome operators are measured repeatedly and
recovery operations are applied~\cite{shor1995scheme,terhal2015quantum,knill1997theory}.
This strategy can reach arbitrary accuracy in principle, but its overhead in physical
qubits is large.
It is therefore of interest to ask how much protection can be supplied by the intrinsic structure of a physical system alone, without syndrome extraction and recovery.

Symmetry-protected topological (SPT) phases provide one possible route toward such protection. As long as the protecting global symmetry is preserved, the ground states of SPT phases exhibit nontrivial nonlocal structures~\cite{Pollmann2012,chen2011classification,ChenXieGuZheng-Cheng, Schuch2011,todadri2015symmetry}.
The corresponding diagnostics include the string order
parameter~\cite{kennedy1992hidden,perez2008string} and the
degeneracy pattern of the entanglement
spectrum~\cite{li2008entanglement,pollmann2010entanglement}.
Under open boundary conditions, this nontrivial SPT order is manifested through symmetry-protected edge modes localized at the boundaries.
These edge modes serve not only as signatures of the phase but also as a candidate code space for quantum information, while the symmetry itself constrains the logical operations that can be implemented within this space.

The canonical solvable example in one dimension is the Affleck-Kennedy-Lieb-Tasaki (AKLT) model~\cite{affleck1987h,affleck1988valence,kennedy1990exact}, which realizes the Haldane phase~\cite{haldane1983nonlinear}.
Each spin-1 is viewed as the symmetric
combination of two virtual spin-$1/2$ degrees of freedom, and virtual spins on
neighboring sites are paired into
singlets~\cite{affleck1987h,affleck1988valence,kennedy1990exact}.
The ground state admits an exact matrix product state (MPS) representation~\cite{fannes1992finitely,perez2006matrix,schollwock2011density,cirac2021matrix},
and it is the zero-energy ground state of a two-body frustration-free parent
Hamiltonian.
For an open chain, one unpaired spin-$1/2$ degree of freedom remains at each end, resulting in a fourfold degenerate ground-state manifold.

A well-established approach to utilizing SPT phases as quantum information resources is measurement-based quantum computation (MBQC)~\cite{raussendorf2001one}. In this framework, the entanglement in the bulk of a quantum state serves as the computational resource, while adaptive local measurements drive the computation.
The one-dimensional spin-1 AKLT chain enables the implementation of arbitrary single-qubit gates in the correlation space~\cite{brennen2008measurement,bartlett2010quantum}, and the
AKLT state on the two-dimensional honeycomb lattice is a universal
resource~\cite{wei2011affleck,wei2018quantum}.
The computational power was later
shown to be a property of an entire SPT phase rather than of isolated
points~\cite{else2012symmetry,stephen2017computational}.
Beyond conventional two-dimensional resource states, one-dimensional architectures supporting universal MBQC have also been proposed~\cite{stephen2024universal}.
On the experimental side, the AKLT state
has been prepared deterministically at constant depth on a quantum
processor~\cite{smith2023deterministic}.

Another approach exploits the robustness provided by the bulk gap and protecting symmetry against symmetry-preserving perturbations.
The edge modes of the one-dimensional Haldane phase can function as a quantum logical wire~\cite{miyake2010quantum}.
Quantum information can be encoded in the degenerate ground space, and logical operations can be implemented using a global field~\cite{bacon2013adiabatic}.
This construction was subsequently extended to general SPT phases by driving the system through a symmetry-preserving phase transition to the trivial phase~\cite{williamson2015symmetry}.
In this setting, a global symmetry operation acts on the MPS virtual level through an intertwining
relation~\cite{perez2008string, Schuch2011}, so that its effect on
the state is confined to the two boundaries.
This is the approach investigated in the present work.

Against this background, three aspects remain to be quantitatively characterized.
The first question concerns the symmetries used to build logical gates.
Existing edge-encoding schemes are formulated on a single chain, so that the available symmetries are internal ones such as spin rotation and time reversal.
A quasi-one-dimensional ladder possesses in addition spatial symmetries, for
instance the exchange of two legs.
Such an operation also admits a virtual
representation, and therefore also induces an operation on the edge subspace, but
it has not been analyzed.

The second concerns the accessibility of the code space to local operators.
The statement that information stored in an SPT phase is invisible to local operators
holds at the level of phase classification, in the sense that no local order
parameter distinguishes distinct SPT
phases~\cite{Pollmann2012,chen2011classification}.
Available robustness results
address whether the phase and its edge degeneracy survive a symmetric local
perturbation~\cite{Pollmann2012,chen2011classification}, and, for open systems,
whether channels obeying a strong symmetry condition preserve one-dimensional SPT
order~\cite{degroot2022symmetry}.
These are statements about the phase.
For encoding, what is missing is a quantitative calculation result that states how much information about the encoded states a given local operator can extract.

The third question concerns the geometry itself.
Although the odd-even structure of spin ladders is well understood in the
condensed matter context~\cite{dagotto1996surprises}, with odd-leg AKLT ladders in the nontrivial Haldane phase and even-leg ladders in the trivial phase~\cite{Pollmann2012,chen2011classification}, these studies address phase
diagrams and bulk properties rather than the encoded subspace.
Increasing the number of legs $M$ in an AKLT ladder modifies two relevant features.
The dimension of the encoded subspace grows as $K=2^{2M}$.
Moreover, the doubled virtual space, on which the transfer matrix acts, extends up to angular momentum $\mathcal L=M$.
For a single chain, the doubled virtual space reaches only $\mathcal L=1$, so that an operator of spherical tensor rank two has no propagation channel at all, and no rank-resolved structure can be observed.

The three-leg ladder provides the minimal setting in which these features are present
simultaneously.
It is the smallest multi-leg AKLT ladder lying in the nontrivial Haldane phase, since the two-leg ladder is trivial.
It is the smallest geometry carrying both the continuous rotation symmetry and a leg exchange that leaves the
middle leg invariant, so that the full on-site symmetry group is
$SO(3)\times\mathbb{Z}_2$.
Its doubled virtual space is the smallest one extending to $\mathcal L=3$, which is what makes a hierarchy of sector-resolved decay lengths observable.
At the same time, the model remains exactly solvable.

In this work we study the three-leg AKLT ladder, using the spin-1 chain and the
two-leg ladder as reference cases. We first construct the exact MPS tensor from
Clebsch-Gordan projectors together with a two-body frustration-free parent
Hamiltonian, identify the full on-site symmetry group and its virtual
representation, and obtain an orthonormal basis of the encoded subspace, which
has dimension $K=64$ and thus carries six logical qubits.
We then demonstrate that the model exhibits a nontrivial Haldane phase
by calculating
$H^2\!\left(SO(3)\times\mathbb{Z}_2,U(1)\right)$, together with string-order parameters and degeneracy patterns of the entanglement spectrum, using the two-leg ladder as a trivial-phase reference.

We first examine the logical gates induced by the on-site symmetry.
The intertwining mechanism itself is standard~\cite{perez2008string,Schuch2011}. The continuous $SO(3)$ factor yields a three-parameter family of logical rotations,
which acts as three parallel copies of the edge operation realized in the spin-1 chain.
The leg exchange $P_{tb}$ factor behaves differently.
Its virtual representation is a swap of two of the three virtual qubits, and the induced
logical gate permutes logical qubits and cannot be written as a tensor product of
single-qubit unitaries.
This gate type has no counterpart in any single-chain AKLT system.
Its scope should be stated precisely.
The gate is a permutation. It generates no entanglement, and it does not render the gate set universal, since a global symmetry operation necessarily acts identically on all virtual qubits of a given boundary.
The point we wish to make is only that the types of symmetry-protected gates available are not fixed by the on-site symmetry group
alone, and that geometry supplies an additional source.

Our second result concerns the accessibility of the encoded subspace by local operators.
We introduce a distinguishability measure $\delta_{\rm edge}$, motivated by the Knill–Laflamme condition~\cite{knill1997theory}.
Using the channel decomposition of the transfer matrix, we derive the spatial decay behavior of $\delta_{\mathrm{edge}}$ with respect to the position of the local probe operator, starting from the definition of the distinguishability measure.
We then verify and analyze this behavior through numerical calculations.
Specifically, this decay is governed by a selection rule raised from the Wigner--Eckart theorem~\cite{edmonds1996angular,tung2013group}: a rank-$\ell$ local operator can couple to the encoded subspace only through the $\mathcal L=\ell$ angular-momentum sector of the transfer matrix, and its decay length is given by the correlation length of this sector $\xi(\mathcal L=\ell)$~\cite{schollwock2011density,perez2006matrix,zauner2015transfer}.
We further identify a finite-size channel that is independent of the probe operator position.
This selection rule exhibits two characteristic limiting cases.
For the spin-1 chain, because the doubled virtual space contains no sectors with $\mathcal L\geq 2$, rank-2 operators become completely unable to distinguish the edge states in the thermodynamic limit.
In contrast, for the three-leg ladder, the three sectors $\mathcal L=1,2,3$ coexist, making the hierarchy $\xi(\mathcal L{=}1)>\xi(\mathcal L{=}2)>\xi(\mathcal L{=}3)$ observable.
The leg-exchange symmetry further splits each sector into even- and odd-parity subsectors, providing an exact explanation for the asymptotic behavior of single-leg operators.
Therefore, the key quantities determining the influence of local probe operators on encoded information are their symmetry labels, including the $SO(3)$ tensor rank and the $P_{tb}$ parity, rather than merely their strength or spatial location.
Numerical calculations confirm the above analytical predictions, with the fitted decay lengths agreeing with the theoretical values for all investigated operators.

The remainder of the paper is organized as follows.
Section~\ref{sec:Model} constructs the MPS tensor, the parent Hamiltonian, the
symmetry group with its virtual representation, and the orthonormal code basis.
Section~\ref{sec:bulk} presents the bulk topological diagnostics.
Section~\ref{sec:Logical gates from on-site symmetries} derives the logical gates generated by the continuous
rotation symmetry and by the leg exchange. Section~\ref{sec:code_parameters} introduces
$\delta_{\rm edge}$, presents its spatial profiles, and interprets them through
the $SO(3)$ selection rule. Section~\ref{sec:conclusion} concludes. The
appendices collect the parent-Hamiltonian projectors, the numerical extraction of
the virtual representations, the group cohomology computation, the derivations of
the string order parameter and of the entanglement spectrum, the verification of
the leg-exchange gate, and the sector decomposition data and channel decomposition.

\section{Model, MPS representation, and code-space construction}
\label{sec:Model}
We study a family of AKLT systems with  legs $M=1,2,3$.
We mainly focus on the three-leg ladder.
The one-dimensional spin-1 AKLT chain and the two-leg ladder are also discussed as reference cases.

Throughout the paper, $N$ denotes the number of unit cells.
For the chain, one unit cell is one site.
For the ladders, one unit cell is one rung.
The dimension of the encoded subspace is denoted by $K$, and the corresponding number of logical qubits is $k = \log_2 K$.

\subsection{Local tensor from Clebsch--Gordan projectors and transfer matrix}
\label{sec:CG and transfer matrix}
A three-leg ladder consists of a top spin-$3/2$, a middle spin-2, and a bottom spin-$3/2$ on each rung.
The local physical Hilbert space is therefore
\begin{equation}
\mathcal{H}_{\text{phys}}^{(3)} \cong \mathbb{C}^4 \otimes \mathbb{C}^5 \otimes \mathbb{C}^4,
\end{equation}
whose physical Hilbert-space dimension is $d_{\text{phys}}= 4 \times 5 \times 4 = 80$.

In the AKLT construction, a  physical space with spin $S$ can be constructed by projecting $2S$ spin-1/2 virtual spaces into the fully symmetric subspace. 
Adjacent virtual spins along vertical direction are paired into singlets and do not participate in the horizontal virtual bonds. Along each horizontal leg, the remaining virtual spins are also connected by singlets in the same manner, leaving three unpaired virtual spin-$\frac{1}{2}$ degrees of freedom at each boundary.
Hence the virtual bond dimension is $D = 2^3 = 8$.
For later reference, the spin-1 AKLT chain has $(D, d_{\text{phys}}) = (2,3)$, while the two-leg ladder has $(D, d_{\text{phys}}) = (4,16)$.

The local physical Hilbert space in a rung is equipped with spin operators $\mathbf S_t$, $\mathbf S_m$, and $\mathbf S_b$, acting on the top, middle, and bottom legs, respectively.
Each spin operator is decomposed into its Cartesian components,
\begin{equation}
\mathbf S_{\mathrm q} = S_{\mathrm q}^{x} \vec x + S_{\mathrm q}^{y}\vec y + S_{\mathrm q}^{z}\vec z, \qquad \mathrm q = t, m, b.
\end{equation}
The eigen state of $z$-component operators are chosen as basis and labeled their own eigenvalues,
\begin{align}
S_{t}^{z} |s_t\rangle = s_t |s_t \rangle,\qquad
S_{m}^{z} |s_m\rangle = s_m |s_m\rangle,\qquad
S_{b}^{z} |s_b\rangle = s_b |s_b\rangle.
\end{align}
For the spin-$\frac{3}{2}$ legs,$s_t, s_b = -\frac{3}{2}, -\frac{1}{2}, \frac{1}{2}, \frac{3}{2}$,
while for the spin-$2$ middle leg,
$s_m = -2, -1, 0, 1, 2$.
The local rung basis are defined by direct product,
\begin{equation}
|s_t, s_m, s_b\rangle = |s_t\rangle \otimes |s_m\rangle \otimes |s_b\rangle.
\end{equation}

Since the three spins on each rung are treated as a physical unit cell in the MPS representation, it is useful to introduce the total spin operator acting on a rung. The total spin operator is defined as
\begin{equation}
S_{\text{rung}}^{\mu} = S_{t}^{\mu}\otimes I_m\otimes I_b+ I_t\otimes S_{m}^{\mu}\otimes I_b + I_t\otimes I_m\otimes S_{b}^{\mu}, \qquad \mu = x, y, z,
\label{eq:s_rung}
\end{equation}
where $S_{t}^{\mu}$, $S_{m}^{\mu}$, and $S_{b}^{\mu}$ are the spin components acting on the top, middle, and bottom legs, respectively.

The local MPS tensor of the three-leg ladder can be constructed from the standard AKLT ingredients as follows~\cite{affleck1987h,affleck1988valence,kennedy1990exact}.
On the top and bottom legs, three virtual spin-$1/2$ degrees of freedom are projected onto the fully symmetric spin-$3/2$ sector.
In the middle, four virtual spin-$1/2$ degrees of freedom are projected onto the fully symmetric spin-2 sector.
We denote these Clebsch--Gordan tensors by $P^{(3/2)}$ and $P^{(2)}$ respectively.

The virtual spins are paired into singlets following the above construction.
We write the singlet tensor as
\begin{equation}
\varepsilon = \begin{pmatrix} 0 & 1 \\ -1 & 0 \end{pmatrix}.
\end{equation}
After contracting the vertical singlets and absorbing the horizontal singlet metrics into the right virtual legs, we obtain a uniform MPS tensor $A_{LR}^{s}$, with shape $A \in \mathbb{R}^{8 \times 8 \times 80}$.
Here $s$ denotes the composite physical index of a single rung,
\begin{equation}
s = (s_t, s_m, s_b),
\end{equation}
where $s_t$ and $s_b$ label the $S_z$ eigenvalues of the spin-$\frac{3}{2}$ top and bottom legs, while $s_m$ labels the $S_z$ eigenvalue of the spin-$2$ middle leg. Explicitly, $s_t, s_b \in \left\{-\frac{3}{2}, -\frac{1}{2}, \frac{1}{2}, \frac{3}{2}\right\}, s_m \in \{-2, -1, 0, 1, 2\}$.
For numerical implementation, the composite physical index can be
mapped to a single integer. We introduce local labels $i_t,i_b=0,1,2,3$ and $i_m=0,1,2,3,4,$ corresponding to the above ordering of the $S^z$ eigenvalues. The
physical index is then encoded as
\begin{equation}
s=20i_t+4i_m+i_b,
\qquad s=0,\ldots,79 .
\end{equation}
Therefore, the tensor has $d_{\mathrm{phys}}=80$ physical components.

Here $L$ and $R$ are compound virtual indices,
\begin{equation}
L = (\alpha_t, \alpha_m, \alpha_b), \qquad R = (\beta_t, \beta_m, \beta_b),
\end{equation}
where each \(\alpha,\beta \in \{0,1\}\) corresponds to the state of a virtual spin-1/2 (\(0\) for \(\downarrow\), \(1\) for \(\uparrow\)). To map the compound indices to single integers we use binary encoding
\begin{equation}
L = 4\alpha_{t}+2\alpha_{m}+\alpha_{b},\qquad R = 4\beta_{t}+2\beta_{m}+\beta_{b}. \label{eq:binary_encoding}
\end{equation}

For any left boundary row vector \(\langle L| \in \mathbb{C}^{1\times D}\) and right boundary column vector \(|R\rangle \in \mathbb{C}^{D\times 1}\), the corresponding MPS is
\begin{equation}
|\psi_{L,R}^{(N)}\rangle = \sum_{s_1,\ldots,s_N} \bigl( \langle L| A^{s_1} A^{s_2} \cdots A^{s_N} |R\rangle \bigr) \; |s_1 s_2 \cdots s_N\rangle. \label{eq:generators}
\end{equation}
The matrix product contracts the virtual indices in the usual way.
When \(\langle L|\) and \(|R\rangle\) run over all possible choices, the set of states \(|\psi_{L,R}^{(N)}\rangle\) spans a linear subspace of the physical Hilbert space, namely, the raw boundary-state family.

The matrix product state constructed above is the exact ground state of a frustration-free parent Hamiltonian constructed by the standard projector method~\cite{affleck1987h,affleck1988valence,kennedy1990exact}.
For a pair of interacting physical spins, we denote their spin quantum numbers by $S_1$ and $S_2$, and the corresponding spin operators by $\mathbf{S}_1$ and $\mathbf{S}_2$, which are chosen from the local spin operators $\mathbf{S}_t$, $\mathbf{S}_m$, and $\mathbf{S}_b$ defined above. If the two physical spins are connected by $\mathsf k$ valence-bond singlets, the maximum allowed total spin is $J_{\mathrm{max}}=S_1+S_2-\mathsf k$.
Therefore, every sector with total spin $J>J_{\mathrm{max}}$ is forbidden and is penalized by the parent Hamiltonian.

The coordination number matches the spin on each leg exactly.
All interactions are therefore nearest two-body,
\begin{equation}
H = \sum_{n=1}^{N-1}\!\big[h^{\mathrm{top}}_{n,n+1}
      +h^{\mathrm{mid}}_{n,n+1}+h^{\mathrm{bot}}_{n,n+1}\big]
   +\sum_{n=1}^{N}\!\big[h^{\mathrm{t\text{-}m}}_{n}
      +h^{\mathrm{b\text{-}m}}_{n}\big],
\end{equation}
with leg terms $h^{\mathrm{top}}=P_{J=3}$,
$h^{\mathrm{mid}}=P_{J=4}$, $h^{\mathrm{bot}}=P_{J=3}$ and rung terms $h^{\mathrm{t\text{-}m}}=h^{\mathrm{b\text{-}m}}=P_{J=7/2}$,
Here $P_J(\mathbf{S}_1,\mathbf{S}_2)$ projects the two-spin Hilbert space onto the total-spin-$J$ sector and can be expressed as a polynomial of $\mathbf{S}_1\cdot\mathbf{S}_2$ (Appendix~\ref{app:parentH}).
Each term annihilates the AKLT state, which is thus a zero-energy ground state.
On a periodic ladder, it is the unique ground state; on an open ladder, the ground space has dimension $K=D^2=64$, precisely the encoded subspace analyzed in the following sections.
The model is expected to be gapped in the thermodynamic limit, as is characteristic of the Haldane
phase~\cite{haldane1983nonlinear,Pollmann2012}.

For later use, for a uniform MPS tensor $A$ associated with one unit cell, the transfer matrix is defined as~\cite{schollwock2011density,perez2006matrix}
\begin{equation}
T = \sum_{s=0}^{d_{\mathrm{phys}}-1} A^{s}\otimes \overline{A^{s}}, \label{eq:transfer_T}
\end{equation}
where \(\overline{A^{s}}\) denotes the element-wise complex conjugate of \(A^{s}\).
This operator acts on the virtual space \(\mathbb{C}^{D}\otimes \mathbb{C}^{D}\).
For an operator \(\hat{C}\) acting on one unit cell,
\begin{equation}
T_{C} = \sum_{s,t=0}^{d_{\mathrm{phys}}-1} C_{st}\; A^{s}\otimes \overline{A^{t}}, \qquad C_{st} = \langle s|\hat{C}|t\rangle. \label{eq:transfer_TO}
\end{equation}

\subsection{Physical symmetries and virtual representations}
\label{sec:2.2}

For any product basis state $\lvert s_t, s_m, s_b \rangle$ on a rung, the total magnetic quantum number is $\mathcal M = s_t + s_m + s_b$.
Since $s_t$ and $s_b$ are half-integers and $s_m$ is an integer, $\mathcal M$ is always an integer. Consequently, a $2\pi$ rotation acts as the identity on every basis state, and hence on the entire physical Hilbert space.
Hence the rotation symmetry group is \(SO(3)\), rather than \(SU(2)\). 
In addition, for \(M=3\) there is a geometric on-site symmetry  \(P_{tb}\), which is the exchange of the top and bottom legs while leaving the middle leg invariant.
The full on-site symmetry group is therefore
\begin{equation}
G = SO(3) \times \mathbb{Z}_2,
\end{equation}
where the \(\mathbb{Z}_2\) factor is generated by \(P_{tb}\).

A physical operation \(g\) acts on a single rung by its corresponding unitary matrices \(u_g\). In the MPS framework, a global symmetry operation is implemented on the virtual level by an invertible matrix \(V(g)\) via the intertwining relation~\cite{perez2008string,Schuch2011}
\begin{equation}
  u_g:A^{s}\to A^{s}_g = \sum_{s'} [u_g]_{ss'} A^{s'} = \alpha_g \, V(g) A^{s} V(g)^{-1},
\label{eq:intertwining}
\end{equation}
where the factor \(\alpha_g = \pm 1\) accounts for the possible global phase in the ray representation of the quantum state, and $V(g)^{-1}$ is the inverse of $V(g)$.

To demonstrate the numerical extraction of $V(g)$, here are two particular $\pi$-rotation elements:
\begin{equation}
\label{eq:gz_gx}
g_z = e^{\,i\pi S_{\text{rung}}^z},
\qquad
g_x = e^{\,i\pi S_{\text{rung}}^x}.
\end{equation}
The virtual representations of these symmetry operations, together with the geometric on-site symmetry operator $P_{\mathrm{tb}}$, are extracted numerically from the null space of the linear system derived from~Eq.\eqref{eq:intertwining} (see Appendix~A).
For the three-leg ladder, the virtual representations corresponding to the physical operations $g_x$, $g_z$, and $P_{\mathrm{tb}}$ are given by
\begin{equation}
V_z = (\sigma^z)^{\otimes 3}, \qquad V_x = (\sigma^x)^{\otimes 3},\qquad V(P_{tb}) = \mathrm{SWAP}_{13},
\label{eq:virtual_reps_ptb}
\end{equation}
where \(\mathrm{SWAP}_{13}\) exchanges the first and third virtual qubits, corresponding to the top and bottom legs.
The matrix \(V(P_{tb}) = \mathrm{SWAP}_{13}\) satisfies \(V(P_{tb})^2 = I\), and \(V_z V_x = - V_x V_z\) in addition.
$V(P_{tb})$ commutes with the entire \(SO(3)\) virtual representation: the \(SO(3)\) rotation acts identically on all three virtual spins, while \(P_{tb}\) merely permutes the first and third factors.
Any operation that acts uniformly on all factors commutes with any permutation of the factors.
The application of this symmetry to logical gate construction is discussed in detail in Sec.~\ref{sec:Logical gates from on-site symmetries}.

\subsection{Raw boundary-state family and orthonormalized code basis}
\label{sec:2.3}

For open boundary condition, each pair of virtual boundary indices $(L, R)$ defines a many-body state $\ket{\psi_{L,R}^{(N)}}$.
These states are generally not orthogonal.
Their overlap matrix defines the Gram matrix
\begin{equation}
\bigl(G_N\bigr)_{(L,R),\,(L',R')} = \braket{\psi_{L,R}^{(N)} | \psi_{L',R'}^{(N)}}.
\end{equation}
$G_N$ is of dimension $D^2\times D^2$ and positive semidefinite by construction.

Via the transfer matrix one has~\cite{schollwock2011density,perez2006matrix}
\begin{equation}
\braket{\psi_{L,R}^{(N)} | \psi_{L',R'}^{(N)}}
= \bigl(T^{N}\bigr)_{(L',L),\,(R,R')},
\label{Eq:21}
\end{equation}
where $T^N$ carries row index $(L',L)$ and column index $(R,R')$.
Defining the reshuffling map $\mathcal{R}$ by $[\mathcal{R}(M)]_{(i,j),(i',j')} = M_{(i',i),(j,j')}$, the Gram matrix is thus
\begin{equation}
\label{eq:reshuffle_def}
G_N = \mathcal{R}(T^N).
\end{equation}

The effective edge state dimension is the rank:
\begin{equation}
K = \operatorname{rank}(G_N).
\end{equation}
For the spin-1 chain one finds $K=4$, while for the three-leg ladder one finds $K=64$.
The corresponding number of logical qubits is
\begin{equation}
k_q = \log_2 K.
\end{equation}

In a finite-length system, due to the exponentially decaying but non-zero interaction between the left and right boundaries, the ground-state boundary states are generally not strictly orthogonal.
Hence, to obtain an orthonormal basis of the encoded subspace, diagonalize the Gram matrix as
\begin{equation}
G_N = V \Lambda V^\top,
\end{equation}
where the columns of $V$ are orthonormal eigenvectors corresponding to the nonzero eigenvalues, and $V^\top$ denotes the transpose of $V$, and
\begin{equation}
\Lambda = \operatorname{diag}(\nu_1,\dots,\nu_K), \qquad \nu_\mu > 0.
\end{equation}
We then define the orthonormalization matrix
\begin{equation}
\mathsf{P} = V \Lambda^{-1/2},
\label{Eq:p}
\end{equation}
and the orthonormalized code basis is
\begin{equation}
\ket{e_\mu} = \sum_{\alpha} \mathsf P_{\alpha\mu} \, \ket{\psi_\alpha^{(N)}},
\end{equation}
where $\alpha$ is a shorthand for the raw boundary label $(L, R)$.
By construction,
\begin{equation}
\mathsf P^\top G_N \mathsf P = I_K,
\label{eq:29}
\end{equation}
where $\mathsf P^\top$ denotes the transpose of $\mathsf P$.
Thus, the nonorthogonal raw edge states are converted into an orthonormal basis of the encoded subspace.

\section{Bulk topological diagnostics}
\label{sec:bulk}

The central question is whether the $M$-leg ladder defined in Sec.~\ref{sec:CG and transfer matrix} realizes a nontrivial SPT phase or a trivial one.

\subsection{SPT classification and the odd--even mechanism}
\label{sec:virtual-symmetry}

As established in Sec.~\ref{sec:2.2}, the full on-site symmetry group of the three-leg ladder is
\begin{equation}
  G = SO(3)\times\mathbb{Z}_2,
  \label{eq:symmetry_group}
\end{equation}
where the continuous factor is generated by the total rung spin operators $S^\mu_{\mathrm{rung}}$ and the $\mathbb{Z}_2$ factor is generated by the top--bottom leg exchange $P_{tb}$.

One-dimensional bosonic SPT phases with on-site symmetry $G$ are
classified by the second group cohomology
$H^{2}(G,U(1))$~\cite{chen2011classification,ChenXieGuZheng-Cheng,Pollmann2012,Schuch2011}.
For a uniform MPS tensor $A^{s}_{LR}$ with on-site symmetry action $u_g$(Eq.~\ref{eq:intertwining}),
\begin{equation}
 A_g^{s} = \sum_{s'}[u_g]_{ss'}\,A^{s'} = \alpha_g\, V(g)\,A^{s}\,V(g)^{-1},
\end{equation}
where $\alpha_g$ is an inessential phase.
The operators $V(g)$ form a projective representation,
\begin{equation}
  V(g_1)\,V(g_2) = \omega(g_1,g_2)\,V(g_1 g_2),\qquad
  \omega:G\times G\to U(1),
  \label{eq:projective_rep}
\end{equation}
whose cohomology class $[\omega]\in H^{2}(G,U(1))$ is the SPT invariant~\cite{chen2011classification,Pollmann2012,ChenXieGuZheng-Cheng}. 
Here $\omega(g_1,g_2)$ is the explicit 2-cocycle function.
In contrast, $[\omega]$ denotes the corresponding cohomology equivalence class modulo coboundaries.
Thus, $\omega(g_1,g_2)$ represents a particular realization of the cohomology class $[\omega]$.

We evaluate $H^{2}(SO(3)\times\mathbb{Z}_2,\,U(1))$ using the K\"unneth decomposition for group cohomology in the Borel sense~\cite{ChenXieGuZheng-Cheng}.
For compact groups with trivial action on the divisible coefficient module $U(1)$,
\begin{equation}
  H^{2}(G_1\!\times\!G_2,\,U(1))
  =
  H^{2}(G_1,U(1))\;\oplus\;
  \bigl[H^{1}(G_1,U(1))\otimes H^{1}(G_2,U(1))\bigr]\;\oplus\;
  H^{2}(G_2,U(1)).
  \label{eq:kunneth}
\end{equation}
The individual terms are (see Appendix~\ref{app:cohomology} for details):
\begin{align}
  H^{2}\bigl(SO(3),U(1)\bigr)
  &\;\cong\;\mathrm{Hom}\bigl(\pi_1(SO(3)),\,U(1)\bigr)
   =\mathrm{Hom}(\mathbb{Z}_2,U(1))=\mathbb{Z}_2,
  \label{eq:H2SO3}\\[4pt]
  H^{1}\bigl(SO(3),U(1)\bigr)
  &\;=\;\mathrm{Hom}\bigl(SO(3),U(1)\bigr)=0,
  \label{eq:H1SO3}\\[4pt]
  H^{2}\bigl(\mathbb{Z}_2,U(1)\bigr)&\;=\;0.
  \label{eq:H2Z2}
\end{align}
Equation~\eqref{eq:H2SO3} reflects the fact that $SO(3)\cong\mathbb{RP}^3$ has
fundamental group $\mathbb{Z}_2$, and projective representations of $SO(3)$ are in bijection with the homomorphisms $\pi_1(SO(3))\to U(1)$~\cite{nash1988topology,chen2012symmetry,ChenXieGuZheng-Cheng}.
Eq.~\eqref{eq:H1SO3} holds because $SO(3)$ is semisimple and therefore admits no nontrivial continuous homomorphism to an abelian group.
Eq.~\eqref{eq:H2Z2} follows from the divisibility of $U(1)$:
every $U(1)$-valued 2-cocycle on $\mathbb{Z}_2$ is a coboundary.
Substituting into Eq.~\eqref{eq:kunneth},
\begin{equation}
  H^{2}\!\bigl(SO(3)\times\mathbb{Z}_2,\;U(1)\bigr)
  =\mathbb{Z}_2\;\oplus\;0\;\oplus\;0
  =\mathbb{Z}_2.
  \label{eq:H2full}
\end{equation}
Therefore there are exactly two phases: the trivial phase and the nontrivial Haldane phase~\cite{haldane1983nonlinear,chen2011classification}.

The vanishing of the mixed term and the $H^{2}(\mathbb{Z}_2)$ term already guarantee on general grounds that the leg-exchange symmetry introduces no new invariant.
It is nevertheless instructive to verify this directly in the AKLT virtual representation.
In the three-leg ladder, the virtual bond space is $(\mathbb{C}^2)^{\otimes 3}$ with basis labeled by the three legs.
The physical leg exchange $P_{tb}$ acts on the virtual level as
\begin{equation}
  V(P_{tb}) = \mathrm{SWAP}_{13},
  \label{eq:Vswap}
\end{equation}
the operator that permutes the first and third tensor factors introduced in Sec.~\ref{sec:2.2}.
This matrix satisfies $V(P_{tb})^{2}=I$ and commutes with the $SO(3)$ virtual operators $V(g)$.
The commutation implies that the 2-cocycle factorizes trivially between the $SO(3)$ and $\mathbb{Z}_2$ sectors:
$\omega\bigl((g_1,a_1),(g_2,a_2)\bigr)=\omega_{SO(3)}(g_1,g_2)$ with no
cross contribution.
In other words, for the AKLT on-site representation the mixed term
$H^{1}(SO(3))\otimes H^{1}(\mathbb{Z}_2)$ is realized trivially, consistent
with the abstract vanishing proved above.
The effective SPT classification therefore reduces to the $SO(3)$ factor alone, as in the single-chain
case~\cite{Pollmann2012,chen2011classification}.

We now determine which $\mathbb{Z}_2$ class is realized as a function of the leg number $M$.
In the MPS construction, the virtual bond space for each rung is the tensor product of $M$ spin-$1/2$ spaces, one
per leg; correspondingly the virtual bond dimension is $D=2^M$.
The on-site $SO(3)$ rotation $\mathcal R(\hat n,\theta)$ is implemented on the
virtual space by
\begin{equation}
  V(g) = u_{1/2}(g)^{\otimes M},\qquad g\in SO(3),
  \label{eq:Vg_tensor}
\end{equation}
where $u_{1/2}(g)$ is the spin-$1/2$ projective representation of
$SO(3)$, i.e.\ the defining representation of $SU(2)$ modulo the
$\mathbb{Z}_2$ center.

It is a standard property of group cohomology that the cocycle class of a
tensor-product representation equals the sum of the individual
classes~\cite{ChenXieGuZheng-Cheng,Schuch2011}.
And $[\omega]$ belongs to $H^2(\mathrm{SO}(3)\times\mathbb{Z}_2, U(1)) = \mathbb{Z}_2$.
Writing $[\omega^{(1)}]=1\in\mathbb{Z}_2$ for the nontrivial class
carried by a single spin-$1/2$, the $M$-fold tensor product carries
\begin{equation}
  [\omega^{(M)}] = M\cdot[\omega^{(1)}] = M \bmod 2
  \;\in\;\mathbb{Z}_2.
  \label{eq:cocycle_additive}
\end{equation}

For the three-leg ladder ($M=3$) studied in this work, $[\omega^{(3)}]=1$ and Eq.~\eqref{eq:cocycle_additive} gives the
anticommutation sign $(-1)^3=-1$.
The system therefore belongs to the Haldane phase.
In the following subsections we confirm this prediction quantitatively via
the string order parameter and the entanglement spectrum.

\subsection{Bulk correlation length and string order parameter}
\label{sec:string-order}

As a supporting nonlocal diagnostic, we consider the string order parameter in the thermodynamic limit.

First, the bulk correlation length is extracted from the two largest eigenvalues of the transfer matrix~\cite{schollwock2011density,perez2006matrix},
\begin{equation}
\label{eq:xi-def}
\xi_{\text{bulk}} = -\frac{1}{\ln |\lambda_1 / \lambda_0|}.
\end{equation}
Numerically, we obtain
\begin{align}
\xi_{\text{bulk}}^{(1)} &= 0.910239 \qquad (\text{spin-1 AKLT chain}), \label{eq:xi1} \\
\xi_{\text{bulk}}^{(2)} &= 1.176425 \qquad (\text{two-leg ladder}), \label{eq:xi2} \\
\xi_{\text{bulk}}^{(3)} &= 1.362981 \qquad (\text{three-leg ladder}). \label{eq:xi3}
\end{align}
Thus all three models are short-range correlated, but their nonlocal string diagnostics are sharply different.

We compare the bulk string order parameter for all three systems.
Following~\cite{kennedy1992hidden,perez2008string}, the string order parameter is defined by
\begin{equation}
\label{eq:string-def}
O_{\text{str}}(m) = \Bigl\langle S_0^z \prod_{n=1}^{m-1} e^{i\pi S_n^z}\, S_m^z \Bigr\rangle.
\end{equation}
For the spin-1 chain, $S_n^z$ is the on-site spin operator.
For the ladders, $S_n^z$ denotes the total $S^z$ on rung $n$, which constitutes one unit cell. Therefore the string of operators in~Eq.(\ref{eq:string-def}) is evaluated as a tensor product over the unit cells.
In the MPS formalism, the string order parameter $O_{\text{str}}(m)$ can be expressed in terms of the modified transfer matrix $T_e$ given in Sec.~\ref{sec:CG and transfer matrix}~Eq.(\ref{eq:transfer_TO}), constructed by inserting the string operator $e^{i\pi S^z}$ into the physical index contraction. Specifically, one obtains
\begin{equation}
\label{eq:string-transfer}
O_{\text{str}}(m) = \frac{\langle l|\, T_{S^z}\, (T_e/\lambda_0)^{m-1}\, T_{S^z}\, |r\rangle}{\lambda_0^2},
\end{equation}
where $|r\rangle$ and $\langle l|$ are the dominant left and right fixed points of the standard transfer matrix, and $\lambda_0$ is its leading eigenvalue.
The derivation is given in Appendix~\ref{sec:appB}.

\begin{figure}
    \centering
    \includegraphics[width=0.75\linewidth]{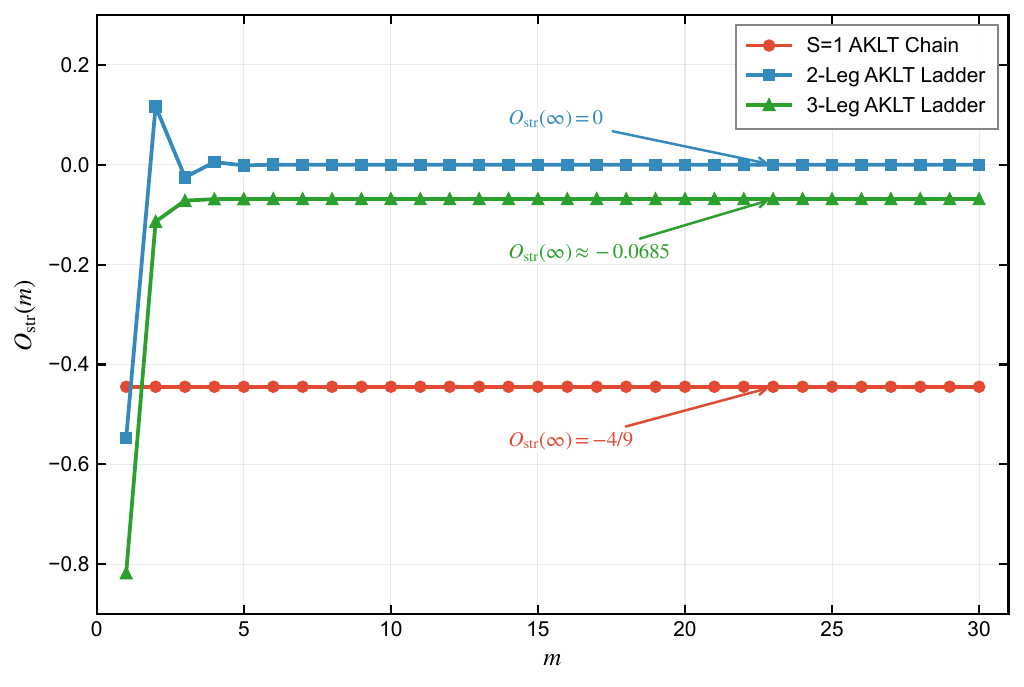}
    \caption{Thermodynamic-limit string order $O_{\text{str}}(m)$ versus $m$ for the chain, two-leg ladder, and three-leg ladder.}
    \label{fig:string-order}
\end{figure}

For the spin-1 AKLT chain, the string order parameter is completely flat as a function of distance and reproduces the standard exact value
\begin{equation}
\label{eq:string-chain}
O_{\text{str}}(m) = -\frac{4}{9}.
\end{equation}
In contrast, the two ladder systems display a pronounced short-distance transient before reaching their large-$m$ asymptotics.
The two-leg ladder shows an alternating sign pattern and decays rapidly to zero. The three-leg ladder instead approaches a finite negative constant,
\begin{equation}
\label{eq:string-3leg}
O_{\text{str}}(\infty) \approx -0.0684710852.
\end{equation}
The resulting behavior is shown in Fig.~\ref{fig:string-order}.

The chain and the three-leg ladder both retain nonzero long-distance string order, while the two-leg ladder does not. This already separates the odd-leg systems from the even-leg one and is consistent with the expected odd--even structure of AKLT ladders.

For the two-leg and three-leg ladders, it is also useful to examine how the string order parameter deviates from its asymptotic value at short distances.

In ladder generalizations, the virtual bond dimension grows and the string transfer matrix $T_e$ acquires a richer eigenvalue spectrum. Moreover, the string operator $e^{i\pi S_{\text{rung}}^z}$, defined as the product of on-site string operators across the rung, no longer reduces to a simple gauge transformation in the enlarged virtual space. The vector $T_{S^z}|r\rangle$ now has nonzero overlap with multiple eigenvectors of $T_e/\lambda_0$.

 Expanding the vector $T_{S^z}|r\rangle$ in the eigenbasis of the normalized string transfer matrix $T_{e}/\lambda_0$, one finds
\begin{equation}
O_{\text{str}}(m) = \sum_i x_i \mu_i^{\,m-1}= O_{\text{str}}(\infty) + \sum_{i\ge 1} x_i \, \mu_i^{\,m-1},
\end{equation}
where $\mu_i$ are the eigenvalues of $T_{e}/\lambda_0$ and the coefficients $x_i$ depend on the overlaps with left and right fixed points (see Appendix~\ref{sec:appB} for a derivation).
In the spin-1 AKLT chain, only the leading mode contributes, so the string order is exactly flat.
In the ladder, several modes contribute, so transient corrections appear at small $m$.
Their decay rates are controlled by the corresponding $\mu_i$.
The transient fluctuations observed numerically are thus a direct manifestation of the multi-component spectral structure of the string transfer matrix in systems with enlarged unit cells.

These results are summarized in Table~\ref{tab:bulk}.

\begin{table}[ht]
\caption{Comparison of the bond dimension $D$, physical dimension
$d_{\text{phys}}$, bulk correlation length $\xi_{\text{bulk}}$,
and asymptotic string order $O_{\text{str}}(\infty)$ for the
three AKLT systems.}
\label{tab:bulk}
\begin{ruledtabular}
\begin{tabular}{lcccc}
 & $D$ & $d_{\text{phys}}$ & $\xi_{\text{bulk}}$ & $O_{\text{str}}(\infty)$ \\
\hline
spin-1 AKLT chain & $2$  & $3$  & $0.910239$ & $-4/9$            \\
two-leg ladder    & $4$  & $16$ & $1.176425$ & $0$               \\
three-leg ladder  & $8$  & $80$ & $1.362981$ & $\approx -0.0685$ \\
\end{tabular}
\end{ruledtabular}
\end{table}

\subsection{Entanglement spectrum and SPT diagnosis}
\label{sec:entanglement}

As a complementary bulk diagnostic, we compute the thermodynamic-limit entanglement spectrum for all three AKLT systems.
The degeneracy pattern of the Schmidt weights provides a direct, MPS-based signature of nontrivial SPT order~\cite{Pollmann2012,chen2011classification,li2008entanglement,pollmann2010entanglement}.

We employ the same transfer-matrix fixed points $\langle l|$ and $|r\rangle$ introduced in Sec.~\ref{sec:string-order}.  
Recall that $\langle l|$ and $|r\rangle$ are the dominant left and right eigenvectors of the transfer
matrix~$T = \sum_{s=0}^{d_{\mathrm{phys}}-1} A^{s}\otimes \overline{A^{s}}$, satisfying
$\langle l | r \rangle = 1$. 
Reshaping these vectors into $D\times D$ matrices yields the Gram matrices
\begin{equation}
\bigl(G_L\bigr)_{(\alpha \beta)} = \langle L_\alpha | L_\beta \rangle, \qquad
\bigl(G_R\bigr)_{(\alpha \beta)} = \langle R_\alpha | R_\beta \rangle,
\end{equation}
where $|L_\alpha\rangle$ and $|R_\alpha\rangle$ are the half-chain states associated with virtual boundary index~$\alpha$ (see Appendix~\ref{sec:appC} for details).

In the thermodynamic limit, the reduced density matrix of either
half-chain is fully determined by these fixed points.  As shown in Appendix~\ref{sec:appC}, the Schmidt weights $\{\lambda_\alpha\}$ are precisely the eigenvalues of the $D\times D$ matrix product~$G_LG_R$:
\begin{equation}\label{eq:ES-formula}
\operatorname{spec}(\rho_L) \;=\; \operatorname{spec}(G_LG_R).
\end{equation}
In practice, after reshaping the fixed-point vectors one symmetrizes $G_L$ and $G_R$ to remove small numerical antisymmetric components, diagonalizes
$G_LG_R$, and normalizes the resulting nonnegative eigenvalues to unit sum.

The entanglement spectra computed from the transfer-matrix fixed points
are as follows.
For the spin-1 AKLT chain, the spectrum consists of a
single twofold-degenerate level: $\lambda_0 = \lambda_1 = 0.500000$.
For the two-leg ladder, the spectrum has a $1+3$
structure: $\lambda_0 = 0.442646$ (non-degenerate) and
$\lambda_{1,2,3} = 0.185785$ (threefold degenerate).
For the three-leg ladder, the spectrum has a $2+2+4$
structure: $\lambda_{0,1} = 0.267123$ (twofold),
$\lambda_{2,3} = 0.084241$ (twofold), and
$\lambda_{4,5,6,7} = 0.074318$ (fourfold).

\begin{figure}
    \centering
    \includegraphics[width=0.75\linewidth]{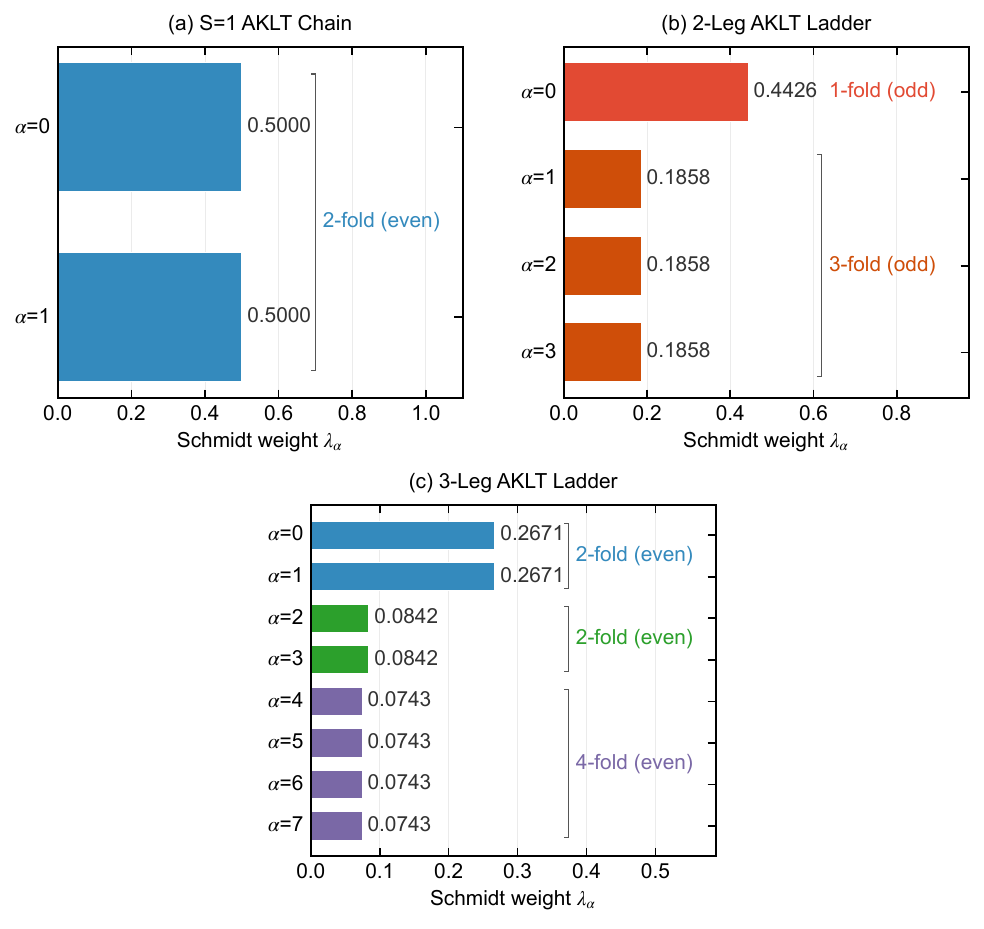}
    \caption{Entanglement spectra of the chain, the two-leg ladder, and the three-leg ladder.}
    \label{fig:entanglement}
\end{figure}

The resulting spectra are shown in Fig.~\ref{fig:entanglement}.

The degeneracy pattern of the entanglement spectrum is a direct consequence of the SPT classification.
We now explain how the $\mathrm{SO}(3)$ symmetry enforces these patterns through the representation theory of the virtual space.

As shown in Sec.~\ref{sec:virtual-symmetry}, the on-site $\mathrm{SO}(3)$ rotation $\mathcal R(\hat{n},\theta)$ is implemented on the virtual space by
\begin{equation}
  V(g) = u_{1/2}(g)^{\otimes M}, \quad g\in \mathrm{SO}(3),
\end{equation}
where $u_{1/2}(g)$ is the spin-$1/2$ projective representation of $\mathrm{SO}(3)$, i.e.\ the defining representation of $\mathrm{SU}(2)$ modulo the $\mathbb{Z}_2$ center.
The cocycle class carried by this $M$-fold tensor product is $[\omega^{(M)}] = M \bmod 2 \in \mathbb{Z}_2 \cong H^2(\mathrm{SO}(3),U(1))$.
For odd $M$ the virtual representation is genuinely projective, placing the system in the nontrivial Haldane phase; for even $M$ it is linear, corresponding to the trivial phase~\cite{chen2011classification,Pollmann2012}.

The link to the entanglement spectrum proceeds as follows. For a uniform MPS tensor $A$, the entanglement spectrum of a half-infinite bipartition equals the spectrum of the product $G_LG_R$ of the left and right fixed-point matrices of the transfer matrix $T = \sum_{s=0}^{d_{\mathrm{phys}}-1} A^{s}\otimes \overline{A^{s}}$. 
We prove in Appendix~\ref{app:commutation} that
\begin{equation}\label{eq:LR-commutation}
  [G_L G_R,\, V(g)] = 0, \quad \forall\, g\in \mathrm{SO}(3).
\end{equation}
The operator $G_LG_R$ therefore commutes with the full virtual $\mathrm{SO}(3)$ action.
Hence by Schur's lemma, $G_LG_R$ is proportional to the identity on each irreducible subspace of $V(g)$.
More precisely, if the virtual space decomposes under $V(g)$ as
\begin{equation}
  (\mathbb{C}^2)^{\otimes M} \cong \bigoplus_{j}\; n_j  \mathbb{C}^{2j+1},
\end{equation}
where $n_j$ is the multiplicity of the spin-$j$ irreducible representation, then $G_L G_R$ takes the block form
\begin{equation}\label{eq:schur-block}
  G_LG_R \big|_{j\text{-sector}} = M_j \otimes I_{2j+1},
\end{equation}
where $M_j$ is an $n_j\times n_j$ non--zero diagonal matrix and $I_{2j+1}$ is the identity on the $(2j+1)$-dimensional irreducible representation~\cite{tung2013group}. Each eigenvalue of $M_j$ thus appears with degeneracy $2j+1$ in the entanglement spectrum.

The SPT diagnostic follows immediately. For odd $M$, the tensor product of an odd number of spin-$1/2$ representations decomposes exclusively into half-integer spin representations, for which $2j+1$ is always even.
All entanglement spectrum degeneracies are therefore forced to be even~\cite{Pollmann2012}.
For even $M$, only integer spin representations appear, and $2j+1$ is odd, so odd degeneracies are symmetry-allowed. The presence or absence of odd degeneracies thus serves as a direct diagnostic of the SPT phase.

The entanglement-spectrum data are fully consistent with the
cohomological prediction of Sec.~\ref{sec:virtual-symmetry}.
Together with the string-order and bulk-correlation-length
results of Sec.~\ref{sec:string-order}, this provides a complete set of
quantitative bulk diagnostics confirming the odd--even stacking rule.

\section{Symmetry-protected logical gates}
\label{sec:Logical gates from on-site symmetries}

\subsection{Continuous SO(3) gate family}
\label{sec:symmetry group}

The discrete generators $g_z$ and $g_x$ studied in Sec.~\ref{sec:2.2} are special elements of the continuous on-site symmetry group SO(3).
In this subsection, we exploit the full continuous symmetry to construct a three-parameter family of logical gates parametrized by an arbitrary rotation $\mathcal{R}(\hat{n},\theta)\in\mathrm{SO}(3)$.

Consider a rotation $\mathcal R(\hat{n},\theta) = e^{i\theta\,\hat{n}\cdot\mathbf{S}_{\mathrm{rung}}}$ acting on the physical Hilbert space of one rung. Because the total rung spin decomposes as $\mathbf {S}_{\mathrm{rung}} = \mathbf {S}_t\otimes I_m\otimes I_b + I_t\otimes\mathbf {S}_m\otimes I_b + I_t\otimes I_m\otimes\mathbf {S}_b$, and the three subsystems occupy the physical space $\mathcal{H}_{\mathrm{phys}}^{(3)}\cong\mathbb{C}^4\otimes\mathbb{C}^5\otimes\mathbb{C}^4$, the physical rotation factorizes as
\begin{equation}
u_g = e^{i\theta\,\hat{n}\cdot\mathbf{S}_t}\otimes e^{i\theta\,\hat{n}\cdot\mathbf{S}_m}\otimes e^{i\theta\,\hat{n}\cdot\mathbf{S}_b}.
\label{eq:phys_rotation_factorize}
\end{equation}

The MPS tensor is constructed from Clebsch--Gordan projections $P^{(3/2)}$ and $P^{(2)}$ that project the symmetric subspaces of virtual spin-$1/2$ degrees of freedom onto the physical spins (Sec.~\ref{sec:CG and transfer matrix}). A fundamental property of CG coefficients is that they intertwine the rotation action between the coupled and uncoupled bases~\cite{tung2013group}.
Applying this property to the top leg ($S=3/2$, $n_S=3$), the middle leg ($S=2$, $n_S=4$), and the bottom leg ($S=3/2$, $n_S=3$), respectively, we obtain the representation of the physical rotation of the entire rung in the virtual space:
\begin{equation}
\begin{split}
&\left[ \mathcal D^{(S_t)}(\mathcal R) \otimes \mathcal D^{(S_m)}(\mathcal R) \otimes \mathcal D^{(S_b)}(\mathcal R) \right] \left( P^{(3/2)} \otimes P^{(2)} \otimes P^{(3/2)} \right)\\ 
&= \left( P^{(3/2)} \otimes P^{(2)} \otimes P^{(3/2)} \right) \left[ \mathcal D^{(1/2)}(\mathcal R) \right]^{\otimes (3+4+3)},
\end{split}
\label{eq:CG_intertwining}
\end{equation}
where $\mathcal D^{(S)}(\mathcal R)$ denotes the spin-$S$ rotation matrix, and $\mathcal D^{(1/2)}(\mathcal R) = e^{i\theta\,\hat{n}\cdot\vec{\sigma}/2}$ is the fundamental spin-$1/2$ rotation.
Define the total projection operator of a rung as
\begin{equation}
P_{\mathrm{rung}}=
P^{(3/2)}\otimes P^{(2)}\otimes P^{(3/2)}.
\end{equation}
To explicitly specify the order of composition of the mappings, the above relation can be abbreviated as
\begin{equation}
u_g\circ P_{\mathrm{rung}} = P_{\mathrm{rung}}\circ
\left[\mathcal D^{(1/2)}(\mathcal R)\right]^{\otimes 10}.
\end{equation}
This equation should be understood as an equality between compositions of linear maps from the tensor-product space of ten virtual spin-$1/2$ degrees of freedom to the physical rung Hilbert space $\mathbb{C}^{4}\otimes\mathbb{C}^{5}\otimes\mathbb{C}^{4}$.
Since $P_{\mathrm{rung}}$ is a non-invertible rectangular projection from the higher-dimensional virtual space to the lower-dimensional physical space, it cannot be rewritten as a similarity transformation of $u_g$. Nevertheless, this does not affect the validity of the above equation, which holds pointwise on its domain of definition.

This intertwining relation, combined with the singlet structure of the vertical and horizontal bonds, which are rotationally invariant, implies that the physical rotation on each leg is pushed through the CG projection onto the corresponding virtual spin-$1/2$ factor.
To see the mechanism of internal cancellation, we perform an explicit tracking of degrees of freedom: the 10 virtual spins of a single rung are decomposed into $3$ (left horizontal singlet) $+2$ (vertical singlet between top and middle) $+2$ (vertical singlet between middle and bottom) $+3$ (right horizontal singlet); between adjacent rungs, the ``right 3'' of the $j$-th rung and the ``left 3'' of the $(j+1)$-th rung form 3 horizontal singlets.
Since any singlet bond is rotationally invariant,
\begin{equation}
\bigl[\mathcal D^{(1/2)}(\mathcal R)\otimes \mathcal D^{(1/2)}(\mathcal R)\bigr] \lvert \text{Singlet} \rangle = \lvert \text{Singlet} \rangle,
\end{equation}
the projection of the rotation operator onto the singlet subspace is exactly the identity operator.
Therefore, all internally paired vertical and horizontal singlets completely absorb the rotation action, leaving only 3 unpaired dangling virtual spins at the leftmost and rightmost boundaries of the entire ladder.

When the physical rotation is decomposed into the action on a single boundary, the lifting from the low-dimensional physical space to the high-dimensional single-sided virtual space involves the non-invertible projection, and therefore introduces a $U(1)$ gauge freedom.
The phase factor $\omega(\mathcal R)$ represents the gauge redundancy of the projective representation, which is retained in the virtual representation of a single-boundary.
Hence, for any $\mathcal R(\hat{n},\theta)\in\mathrm{SO}(3)$, the virtual representation of a single-boundary takes the form
\begin{equation}
V\bigl(\mathcal R(\hat{n},\theta)\bigr) = \omega(\mathcal R)\cdot \bigl(e^{i\theta\,\hat{n}\cdot\vec{\sigma}/2}\bigr)^{\otimes 3},
\label{eq:continuous_V}
\end{equation}
where $\omega(\mathcal R) = \pm 1$ is the projective phase shown in Sec.~\ref{sec:virtual-symmetry}, and the three tensor factors act on the top, middle, and bottom virtual qubits respectively.

The form of the single-boundary virtual representation $V\bigl(\mathcal R(\hat{n},\theta)\bigr)$ has been established in Eq.~(\ref{eq:continuous_V}).
We now derive the logical gate induced by a global rotation.
Consider applying the same rotation $\mathcal R(\hat{n},\theta)$ simultaneously to all physical sites.
According to Eq.~(\ref{eq:intertwining}),
for an open-boundary MPS of length \(\mathsf L\), the virtual transformation in the overall wavefunction can be expressed as
\begin{equation}
\prod_{i=1}^{\mathsf L} \left( \alpha_g \, V A^{s_i} V^{-1} \right) = \alpha_g^\mathsf{L}  V A^{s_1} V^{-1} \, V A^{s_2} V^{-1} \cdots V A^{s_\mathsf L} V^{-1}
= \alpha_g^\mathsf{L} V A^{s_1} A^{s_2} \cdots A^{s_\mathsf L} V^{-1},
\end{equation}
where all \(V^{-1}V\) factors between adjacent tensors in the bulk cancel completely, and the factor $\alpha_g^\mathsf L$ contributes only a global phase to the many-body wave function and therefore does not affect the induced logical operation.

Therefore, the global physical rotation leaves no nontrivial action in the interior of the system; it only produces virtual boundary transformations at the two open boundaries:
\begin{equation}
V_L(\mathcal R) = V\mathcal (\mathcal R), \qquad V_R(\mathcal R) = V(\mathcal R)^{-1}.
\end{equation}
For the AKLT ladder, the virtual boundary degrees of freedom are mapped to the boundary encoding subspaces in the physical Hilbert space via the injective MPS boundary maps. Consequently, the above virtual boundary representation induces equivalent logical operations in the physical encoding space.
It is important to emphasize that the virtual boundary space and the physical encoding space are not the same space, but are related by a one-to-one correspondence established through the MPS construction~\cite{perez2006matrix,Schuch2011}.
Hence, the following logical gates are generated by the virtual boundary representation and correspond to their respective actions in the physical encoding space.

Thus, according to Eq.~(\ref{eq:continuous_V}), the virtual logical operations on the left and right boundaries are respectively
\begin{equation}
\mathcal O_L(\mathcal R) = \omega(\mathcal R) \bigl(e^{i\theta\,\hat{n}\cdot\vec{\sigma}/2}\bigr)^{\otimes 3}, \qquad
\mathcal O_R(\mathcal R) = \omega(\mathcal R)^{-1} \bigl(e^{-i\theta\,\hat{n}\cdot\vec{\sigma}/2}\bigr)^{\otimes 3}.
\end{equation}
Define
\begin{equation}
\mathcal U(\mathcal R) = \bigl(e^{i\theta\,\hat{n}\cdot\vec{\sigma}/2}\bigr)^{\otimes 3},
\end{equation}
then the virtual representation on a single boundary can be written as
\begin{equation}
V(\mathcal R) = \omega(\mathcal R)\, \mathcal U(\mathcal R).
\end{equation}
Therefore, the left and right boundary actions induced by the global rotation satisfy
\begin{equation}
V_L(\mathcal R) \otimes V_R(\mathcal R)
= \bigl[\omega(\mathcal R) \mathcal U(\mathcal R)\bigr] \otimes \bigl[\omega(\mathcal R)^{-1} \mathcal U(\mathcal R)^{-1}\bigr]
= \mathcal U(\mathcal R) \otimes \mathcal U(\mathcal R)^{-1}.
\end{equation}
Consequently, the logical operation implemented by the global rotation in the encoding space can be written as
\begin{equation}
\mathcal O_{\text{logical}}(\mathcal R) \cong
\bigl(e^{i\theta\,\hat{n}\cdot\vec{\sigma}/2}\bigr)^{\otimes 3}
\otimes
\bigl(e^{-i\theta\,\hat{n}\cdot\vec{\sigma}/2}\bigr)^{\otimes 3},
\end{equation}
where the symbol \(\cong\) indicates that this expression is the equivalent logical action obtained under the MPS correspondence between the virtual boundary representation and the physical boundary encoding space.

At $\theta=\pi$, the continuous family reduces to the discrete gates:
\begin{equation}
\mathcal O\bigl(\mathcal R(\hat{z},\pi)\bigr) = \mathcal O(g_z), \qquad \mathcal O\bigl(\mathcal R(\hat{x},\pi)\bigr) = \mathcal O(g_x),
\end{equation}
which decompose as $\mathcal O(g_z)=\sigma_z^{\otimes 6}$ and $\mathcal O(g_x)=\sigma_x^{\otimes 6}$. Numerically, each has eigenvalue $+1$ with multiplicity~32 and $-1$ with multiplicity~32, reflecting the balanced $\mathbb{Z}_2$ action on the $K=64$-dimensional physical boundary encoding space.

Specific choices of angle yield gates of independent interest.
At $\theta=\pi/2$, the $z$-rotation produces the phase gate $\mathsf S = \mathcal R_z(\pi/2)$ on each left-boundary qubit and $\mathsf S^{\dagger}$ on each right-boundary qubit, while the $x$-rotation produces $\mathcal R_x(\pi/2)$ and $\mathcal R_x(-\pi/2)$ respectively.
Both are Clifford gates~\cite{nielsen2010quantum}.
At $\theta = \pi/4$, the $z$-rotation produces $\mathsf T = \mathcal R_z(\pi/4)$, which is a non-Clifford gate. Since both Clifford and non-Clifford single-qubit rotations are accessible within this gate family, the continuous SO(3) symmetry provides a richer set of deterministic logical operations than the discrete $\pi$-rotations alone.

\subsection{Logical gate from the discrete geometric reflection}
\label{sec:logical gates}

In addition to the continuous SO(3) symmetry, the three-leg AKLT ladder possesses a discrete geometric symmetry generated by the top--bottom leg exchange $P_{tb}$. As established in Sec.~\ref{sec:2.2}, this operation swaps the spin-3/2 degrees of freedom on the top and bottom legs while leaving the middle spin-2 leg invariant. Since $P_{tb}^2 = I$ and $[P_{tb}, \mathcal R(\hat{n},\theta)] = 0$ for all rotations, it generates a $\mathbb{Z}_2$ factor, and thus produces the full on-site symmetry group $G = \mathrm{SO}(3)\times\mathbb{Z}_2$.

This geometric leg-exchange symmetry is a feature unique to multi-leg ladder geometries and has no analogue in any single-chain AKLT system. The analysis of its virtual representation and the resulting logical gate constitutes one of the new results of this work.

Using the SVD null-space extraction procedure (Appendix~\ref{sec:appD}), we find that the virtual representation $V(P_{tb})$ acts on the three-qubit virtual basis $|\alpha_t\,\alpha_m\,\alpha_b\rangle$ as
\begin{equation}
V(P_{tb})\,|\alpha_t\,\alpha_m\,\alpha_b\rangle = |\alpha_b\,\alpha_m\,\alpha_t\rangle,
\label{eq:vptb_action}
\end{equation}
i.e.\ it exchanges the first and third virtual qubits while leaving the middle qubit untouched.
This is the permutation operator $\mathrm{SWAP}_{13}$ on a three-qubit register.

Following the same transfer-matrix mechanism as in Sec.~\ref{sec:symmetry group}, the logical gate of boundary encoding subspaces is
\begin{equation}
\mathcal O(P_{tb}) = V(P_{tb}) \otimes V(P_{tb})^{-1}.
\label{eq:optb_formula}
\end{equation}
Since $\mathrm{SWAP}_{13}$ is a real symmetric involution, which is $\mathrm{SWAP}_{13}^{-1} = \mathrm{SWAP}_{13}$, this simplifies to
\begin{equation}
\mathcal O(P_{tb}) = \mathrm{SWAP}_{13} \otimes \mathrm{SWAP}_{13},
\label{eq:optb_swap}
\end{equation}
satisfying $\mathcal O(P_{tb})^2 = I_K$ with $K = 64$.
Its eigenvalue spectrum consists of $+1$ with multiplicity~40 and $-1$ with multiplicity~24 (see Appendix~\ref{app:40+24} for details), in contrast to the balanced $32{+}32$ split of the rotation gates at $\theta=\pi$.

The gate $\mathcal O(P_{tb})$ cannot be decomposed into a direct product of single-qubit operations. In the six-qubit encoding where logical qubits 1--3 reside on the left boundary and qubits 4--6 on the right, $\mathcal O(P_{tb})$ simultaneously permutes qubits $1\leftrightarrow 3$ and qubits $4\leftrightarrow 6$.
Such a permutation cannot be expressed as a product $U_1\otimes U_2\otimes\cdots\otimes U_6$ of single-qubit unitaries (Appendix~\ref{app:single-qubit gates}). By contrast, the rotation gates at $\theta=\pi$ decompose as $\mathcal O(g_z)=\sigma_z^{\otimes 6}$ and $\mathcal O(g_x)=\sigma_x^{\otimes 6}$, which are manifestly products of single-qubit Pauli operators. 
Thus, it is a genuinely multi-qubit operation arising from a geometric leg-exchange symmetry.
This demonstrates that in the AKLT system, geometric symmetries can give rise to a multi-qubit permutation logical gate.

Together with the full continuous family $\{\mathcal O(R)\}_{R\in\mathrm{SO}(3)}$, the gate $\mathcal O(P_{tb})$ enriches the set of logical gates that can be derived from the full on-site symmetry group $G = SO(3) \times \mathbb{Z}_2$. 
But this gate set is not universal, consistent with the general understanding that on-site symmetries of a one-dimensional SPT phase cannot yield a universal gate set through the MPS intertwining mechanism alone~\cite{ stephen2017computational, else2012symmetry}.

\section{Local probes of the edge modes and SO(3) selection rules}
\label{sec:code_parameters}

In Sec.~\ref{sec:Logical gates from on-site symmetries}, we studied the logical gates induced by global symmetry operations using the virtual boundary representation of the MPS.
In practice, the actual quantum encoding takes place in the physical boundary encoding subspace, which is isomorphic to the virtual edge space.
The degenerate edge subspace emerges from the nontrivial SPT phase under open 
boundary conditions.
It is therefore instructive to investigate the accessibility of this subspace to local operators: namely, which local probes can extract information about the edge degrees of freedom, and how their probing capability decays as the probe moves into the bulk.

\subsection{Distinguishability measure and profiles}
\label{sec:kl_profiles}

\begin{figure}
\centering
\includegraphics[width=0.9\linewidth]{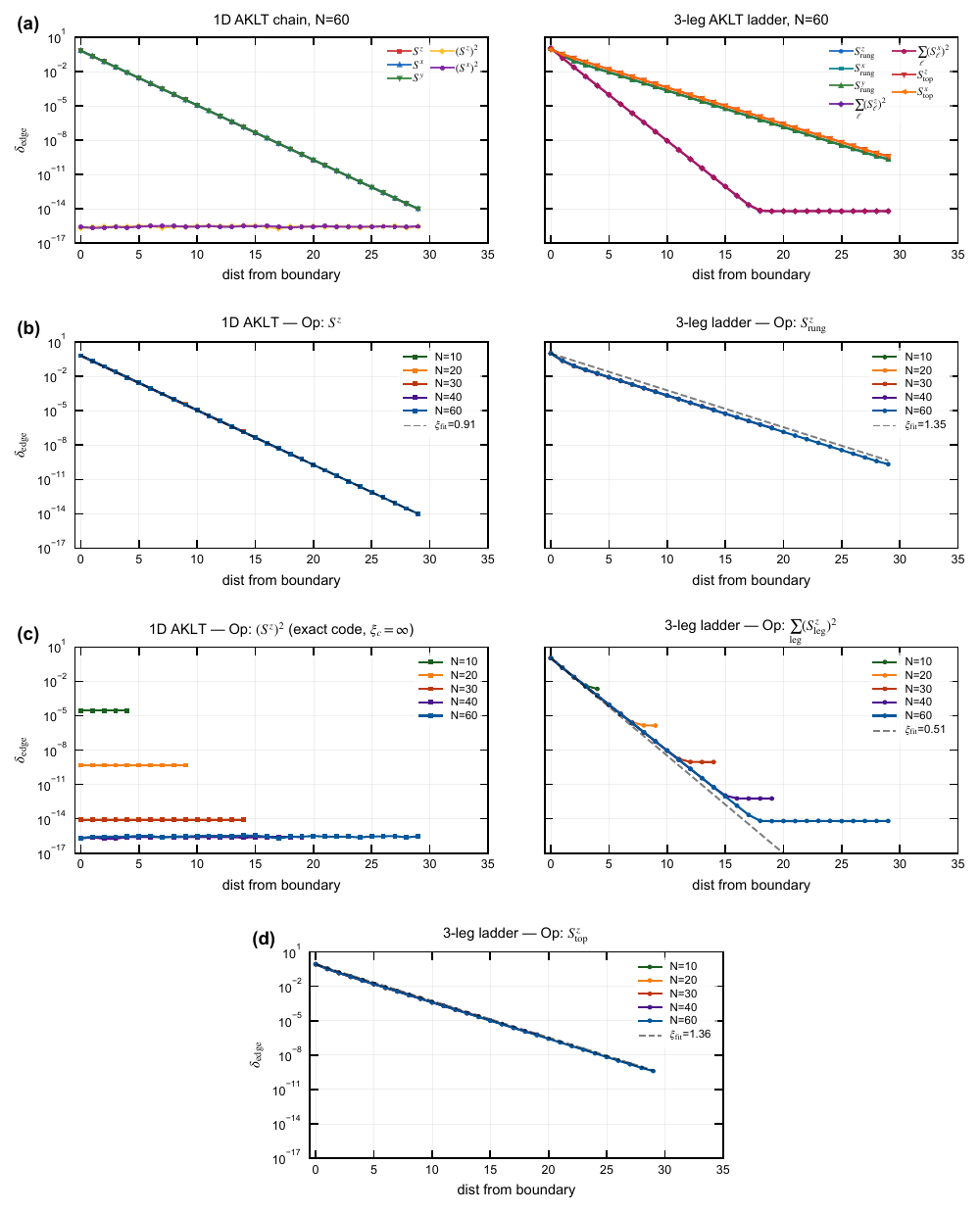}
\caption{(a) Edge-distinguishability $\delta_{\mathrm{edge}}$ versus
distance from the boundary for all operators at fixed $N=60$, for the chain
(left) and the three-leg ladder (right). (b) Size dependence of
$\delta_{\mathrm{edge}}$ for the rank-1 operator $S^z$ in the chain and
$S^z_{\mathrm{rung}}$ in the three-leg ladder. (c) Size dependence for the
rank-2 operator $(S^z)^2$ in the chain and $\sum_\text{leg}(S^z_\text{leg})^2$ in the
three-leg ladder. (d) Size dependence for the single-leg operator
$S^z_{\mathrm{top}}$ in the three-leg ladder.}
\label{fig:kl}
\end{figure}

For a single-site operator $\hat{F}$ acting on unit cell $k$, its matrix
in the raw boundary basis is
\begin{equation}
\label{eq:F_raw}
(F_{\mathrm{raw}})_{\beta\alpha}
  = \braket{\psi_\beta^{(N)} | \hat{F} | \psi_\alpha^{(N)}},
\end{equation}
and its projection onto the orthonormalized edge basis established in Sec.~\ref{sec:2.3} is
\begin{equation}
\label{eq:Oedge_def}
F_{\mathrm{edge}} = \mathsf P^\top\, F_{\mathrm{raw}}\, \mathsf P,
\end{equation}
where $\mathsf P^\top$ denotes the transpose of $\mathsf P$. We characterize how strongly $\hat{F}$ acts on the edge subspace through
\begin{equation}
\label{eq:c_deltaedge}
c(F) = \frac{\mathrm{Tr}(F_{\mathrm{edge}})}{K}, \qquad
\delta_{\mathrm{edge}}(F)
   = \frac{1}{\sqrt{K}}
     \bigl\lVert F_{\mathrm{edge}} - c(F)\, I_K \bigr\rVert_\mathrm F,
\end{equation}
where $\|\cdot\|_\mathrm F$ denotes the Frobenius norm.
The factor $1/\sqrt{K}$ makes $\delta_{\mathrm{edge}}$ a root-mean-square
matrix element rather than a sum over all $K^2$ of them, so that the
measure does not grow with the code dimension.
Equivalently, in the normalized Hilbert--Schmidt inner product
$\langle X,Y\rangle = \mathrm{Tr}(X^\dagger Y)/K$, for which $\|I_K\|=1$,
one has the orthogonal decomposition
\begin{equation}
\bigl\lVert F_{\mathrm{edge}} \bigr\rVert^{2}
  = c(F)^{2} + \delta_{\mathrm{edge}}(F)^{2},
\label{eq:pythagoras}
\end{equation}
so that $c(F)$ and $\delta_{\mathrm{edge}}(F)$ are the scalar and the
distinguishable part of $F_{\mathrm{edge}}$ on the same footing.
The quantity $\delta_{\mathrm{edge}}(F)$ is the distinguishability measure.
It vanishes if and only if $\hat{F}$ acts as a pure scalar on the edge
subspace, that is, when the operator cannot distinguish any two edge states
and carries no accessible edge information.
A large $\delta_{\mathrm{edge}}$ means the operator resolves the edge states
strongly.
$\delta_{\mathrm{edge}}$ is normalized by the code dimension but not by the norm of $\hat F$. 
Absolute heights in Fig.~\ref{fig:kl} therefore still reflect the operator scale,
while the decay rates are intrinsic to the code.
According to the Knill--Laflamme conditions~\cite{knill1997theory}, When considering the single-site operator $\hat{F}$ as an error, the operators with a large $\delta_{\text{edge}}$ strongly violate this condition and resolve the encoded states.

The boundary is a special position.  In an open-boundary MPS the physical
operator at $k=0$ acts directly on the left virtual index without any
attenuation by the transfer matrix, and the same holds for the rightmost
site $k=N-1$.  A boundary probe therefore reaches the edge subspace with
full strength, independently of the system size $N$.

In the MPS framework, the raw matrix $F_{\mathrm{raw}}$ for an operator on
site $k$ is obtained by reshuffling $T^k T_OT^{N-1-k}$,
where $T$ and $T_O$ are defined in
Eqs.~\eqref{eq:transfer_T}--\eqref{eq:transfer_TO}.
Here the Kronecker product in
Eqs.~\eqref{eq:transfer_T}--\eqref{eq:transfer_TO} joins the ket and the
bra layer of a single unit cell, whereas the matrix product contracts the
virtual bonds between neighboring cells.
Because the transfer matrix is the same at every site, the profile $\delta_{\mathrm{edge}}(k)$ is symmetric about the chain center.
We therefore plot $\delta_{\mathrm{edge}}$ against the distance to the nearest
boundary,
\begin{equation}
\mathrm{dist}(k) = \min(k,\, N-1-k).
\label{eq:dist_def}
\end{equation}
The profile is strongly nonuniform. It is largest at the boundary and
falls exponentially toward the bulk, so a local probe acting deep inside
the chain is exponentially blind to the edge information.

For the spin-1 chain, we evaluate $\delta_{\mathrm{edge}}$ for the five
single-site operators $S^z$, $S^x$, $S^y$, $(S^z)^2$ and $(S^x)^2$.
For the three-leg ladder, we evaluate the total rung operators $S^z_{\mathrm{rung}}$,
$S^x_{\mathrm{rung}}$,$S^y_{\mathrm{rung}}$, the summed quadratic operators
$\sum_\text{leg} (S^z_\text{leg})^2$ and $\sum_\text{leg} (S^x_\text{leg})^2$ with $\text{leg}$ running over the top, middle, and bottom legs, and the single-leg operators $S^z_{\mathrm{top}}$ and $S^x_{\mathrm{top}}$.
Specifically, $\sum_\text{leg} (S^z_\text{leg})^2= (S^z_t)^2\otimes I_m\otimes I_b +I_t \otimes (S^z_m)^2 \otimes I_b +I_t \otimes I_m \otimes (S^z_b)^2$.

Figure~\ref{fig:kl}(a) shows $\delta_{\mathrm{edge}}$ versus
$\mathrm{dist}$ for all operators at fixed $N=60$. Three features stand
out.

(i) Within each model, operators of the same SO(3) tensor rank produce
identical profiles. In the spin-1 chain, $S^x$, $S^y$, and $S^z$ coincide,
and so do $(S^z)^2$ and $(S^x)^2$. In the three-leg ladder,
$S^z_{\mathrm{rung}}$, $S^x_{\mathrm{rung}}$, $S^y_{\mathrm{rung}}$
coincide; $\sum_\text{leg}(S^z_\text{leg})^2$ and $\sum_\text{leg}(S^x_\text{leg})^2$ coincide; and
$S^z_{\mathrm{top}}$ and $S^x_{\mathrm{top}}$ coincide.

(ii) In the spin-1 chain, the rank-2 operators $(S^z)^2$ and $(S^x)^2$ give
$\delta_{\mathrm{edge}}\approx 3\times 10^{-16}$ at every position.
The line is flat at double-precision machine level. These operators carry no
edge information at all in the thermodynamic limit.

(iii) In the three-leg ladder, $\sum_\text{leg}(S^z_\text{leg})^2$ and $\sum_\text{leg}(S^x_\text{leg})^2$ show a two-stage
profile.
It first decays exponentially and then levels off onto a plateau,
which at $N=60$ sits at $\delta_{\mathrm{edge}}\approx 6.5\times10^{-15}$.

Figure~\ref{fig:kl}(b)--(d) show the system-size dependence for
representative operators. In every case, the profiles from different $N$
collapse onto a single exponential envelope in the near-boundary region.
This confirms that the decay rate is a bulk property fixed by the
transfer-matrix spectrum, not a finite-size effect.
Figure~\ref{fig:kl}(c) additionally shows that the plateau of the rank-2
probes is not a numerical artifact.  Its height is independent of position
but decreases as $N$ grows, whereas a double-precision floor would sit at
the same level for every $N$.  

To extract the decay length $\xi_{\mathrm{fit}}$, we fit
$\delta_{\mathrm{edge}}(\mathrm{dist})$ over the left half of the chain,
$\mathrm{dist} = 0, \ldots, \lfloor N/2 \rfloor$, at $N=60$. For operators
whose profile decays monotonically, we take the logarithm of
$\delta_{\mathrm{edge}}(\mathrm{dist}) =  \mathrm  A\,e^{-\mathrm{dist}/\xi}$ and perform ordinary
least-squares regression of $\ln\delta_{\mathrm{edge}}$ against
$\mathrm{dist}$, so that $\xi_{\mathrm{fit}} = -1/\mathrm{slope}$.

For the summed quadratic operators $\sum_{\text{leg}}(S^z_\text{leg})^2$ and
$\sum_\text{leg}(S^x_\text{leg})^2$ of the three-leg ladder the decay is fast, and the profile levels off after about eighteen rungs.
A pure exponential fit over
the full half-chain would be contaminated by this plateau and would
overestimate $\xi$. We therefore use the three-parameter form
\begin{equation}
\delta_{\mathrm{edge}}(\mathrm{dist}) =  \mathrm  A\,e^{-\mathrm{dist}/\xi} + \mathrm C,
\label{eq:3param_fit}
\end{equation}
where $ \mathrm  C$ absorbs the plateau, and fit it by nonlinear least squares. For
the chain $(S^z)^2$, $\delta_{\mathrm{edge}} \approx 3\times10^{-16}$ at all
positions, so no decay can be fitted and the effective $\xi=\infty$. The
numerical results are collected in Table~\ref{tab:fits}.

\begin{table}[ht]
\caption{Fitted decay lengths $\xi_{\mathrm{fit}}$ of the
distinguishability profiles at $N=60$.}
\label{tab:fits}
\begin{ruledtabular}
\begin{tabular}{llcc}
System & Operator & $\xi_{\mathrm{fit}}$ & $\mathrm R^2$ \\
\hline
1D chain & $S^z, S^x, S^y$ & $0.9105$ & $0.999999$ \\
1D chain & $(S^z)^2, (S^x)^2$ & $\infty$ & --- \\
3-leg ladder & $S^z_{\mathrm{rung}}, S^x_{\mathrm{rung}},
S^y_{\mathrm{rung}}$ & $1.3459$ & $0.999082$ \\
3-leg ladder & $\sum_{\text{leg}}(S^z_\text{leg})^2,\;\sum_\text{leg}(S^x_{\text{leg}})^2$ &
$0.5090$ & $0.999996$ \\
3-leg ladder & $S^z_{\mathrm{top}}, S^x_{\mathrm{top}}$ & $1.3585$ &
$0.999915$ \\
\end{tabular}
\end{ruledtabular}
\end{table}

\subsection{Interpretation via SO(3) selection rules}
\label{sec:selection_rules}

The operator groupings, the flat profile of $(S^z)^2$ in the
chain, and the two-stage profile of $\sum_{\text{leg}}(S^z_\text{leg})^2$ in the ladder all follow from the SO(3) sector decomposition of the transfer matrix.
We outline the argument here and give the full spectral
data in Appendix~\ref{app:sectors}.

The transfer matrix $T = \sum_{s=0}^{d_{\mathrm{phys}}-1} A^{s}\otimes \overline{A^{s}}$ acts on the doubled
virtual space $\mathbb{C}^D\otimes\mathbb{C}^D$. Because the AKLT tensor
satisfies the intertwining relation, $T$ commutes with the representation
$V(g)\otimes V(g)$ for all $g\in\mathrm{SO}(3)$. Although $V(g)$ is
projective, the doubled representation $V(g)\otimes V(g)$ is a genuine
linear representation. Its irreducible components are therefore labeled by
integer angular momentum $\mathcal L$. In practice, the block decomposition is
implemented via the adjoint Casimir $\mathbf{J}^2_{\mathrm{adj}}$
constructed from the virtual-space spin generators
(Appendix~\ref{app:sectors}, Eq.~\eqref{eq:J2adj}), whose eigenspaces with
eigenvalue $\mathcal L(\mathcal L+1)$ define the $\mathcal L$-sectors. By Schur's lemma $T$ is
block-diagonal in these sectors, and each eigenvalue carries a $(2\mathcal L+1)$-fold
degeneracy.

The leading eigenvalue $\lambda_0$ sits in the $\mathcal L=0$ sector, which contains
the dominant fixed point $\langle l|r\rangle = 1$. The leading eigenvalue
ratio $|\lambda_L/\lambda_0|$ in each sector defines a sector-dependent
correlation length~\cite{schollwock2011density,zauner2015transfer,perez2006matrix},
\begin{equation}
\xi(\mathcal L) = -\frac{1}{\ln|\lambda_\mathcal L / \lambda_0|}.
\label{eq:xi_L}
\end{equation}
The numerical values are listed in Table~\ref{tab:sectors}.

\begin{table}[ht]
\caption{Sector-dependent decay length $\xi(\mathcal L)$ for each angular-momentum
sector, compared with the fitted values of Table~\ref{tab:fits}.}
\label{tab:sectors}
\begin{ruledtabular}
\begin{tabular}{lccc}
System & Sector $\mathcal L$ & $\xi_{\mathrm{theory}}$ & $\xi_{\mathrm{fit}}$ \\
\hline
1D chain & $1$ & $0.9102$ & $0.9105$ \\
\hline
3-leg ladder & $1$ & $1.3630$ & $1.3459$ \\
3-leg ladder & $2$ & $0.5418$ & $0.5090$ \\
3-leg ladder & $3$ & $0.2960$ & --- \\
\end{tabular}
\end{ruledtabular}
\end{table}

The $\mathcal L=3$ sector of the ladder decays within a single lattice spacing,
$\xi_{\mathrm{theory}}<0.3$, so no exponential regime is resolvable and no
fit is attempted. For both models the leading correlation length coincides
with the $\mathcal L=1$ sector, $\xi_{\mathrm{bulk}}=\xi(\mathcal L=1)$. This is consistent with
\begin{equation}
\xi_{\mathrm{bulk}}
= \max\!\bigl(\xi_0^{\,\mathrm{sub}},\; \xi(\mathcal L=1),\; \xi(\mathcal L=2),\; \xi(\mathcal L=3)\bigr),
\label{eq:xi_bulk_max}
\end{equation}
where $\xi_0^{\,\mathrm{sub}}$ is the correlation length of the sub-leading
eigenvalue in the $\mathcal L=0$ sector.
For the three-leg ladder that eigenvalue lies in the $P_{tb}$-even
subsector and gives $\xi_0^{\,\mathrm{sub}} = -1/\ln(0.1141)=0.4607$, which is not the
$0.3725$ of the $(\mathcal L{=}0,-)$ subsector listed in
Table~\ref{tab:sector_full}.
It confirms that spin--spin correlations
$\langle \mathbf{S}_i \cdot \mathbf{S}_j \rangle$ are the slowest-decaying
channel.
The values $\xi_{\mathrm{bulk}} = 0.9102$ for the chain and
$1.3630$ for the three-leg ladder reproduce those obtained from the full
transfer matrix in Eqs.~\eqref{eq:xi1},\eqref{eq:xi3}, which is an independent check
of the sector decomposition.

By a selection rule raised from the Wigner--Eckart
theorem~\cite{edmonds1996angular,tung2013group}, an operator of
spherical-tensor rank $\ell$ couples the $\mathcal L=0$ sector only to the $\mathcal L=\ell$ sector.
Its distinguishability profile therefore inherits the decay rate of that sector,
\begin{equation}
\delta_{\mathrm{edge}}(\mathrm{dist}) \;\simeq\;
\mathrm A\left|\frac{\lambda_{\mathcal L=\ell}}{\lambda_0}\right|^{\mathrm{dist}}
= \mathrm A\,e^{-\mathrm{dist}/\xi(\mathcal L=\ell)},
\label{eq:dKL_sector}
\end{equation}
where $\lambda_{\mathcal L=\ell}$ is the largest-modulus eigenvalue of that
sector whose reduced matrix element does not vanish.
Appendix~\ref{app:Channel decomposition and derivation} derives Eq.~\eqref{eq:dKL_sector} from an exact
channel decomposition of $F_{\mathrm{raw}}$, and gives its regime of
validity, $1\ll\mathrm{dist}\ll N$.
This is the precise sense in which the SO(3) tensor rank of a probe fixes
how fast it can reach the edge information. We now show how the rule
explains each observation.

One could identify that the plateau in Figure~\ref{fig:kl}(c), which decreases with growing $N$. It comes from the contribution of those channels in which both transfer-matrix factors are non-leading. 
Such a channel needs both boundaries at finite distance, and it is suppressed by the system size, while the decay rate is bounded by $|\lambda_1/\lambda_0|^{\,N-1}$, and not by the distance to the nearest boundary; at $N=20$ it is constant to four
digits across the whole half chain. 
It becomes visible once the rank-2
signal has decayed below it, which for $\sum_\text{leg}(S^z_\text{leg})^2$ happens
after about nine rungs at $N=20$ in Table~\ref{tab:H2}. 
But the flat line at $\sim10^{-15}$ in
the $N=60$ curve is the machine floor.

(i) \emph{Identical profiles for same-rank operators.} The components $S^x$, $S^y$, $S^z$ are the three magnetic components ($q=-1,0,+1$) of a rank-1 spherical tensor.
The selection rule guarantees that all components of a given irreducible tensor share the same reduced matrix element.
Since the edge subspace inherits the full SO(3) symmetry, the projected operators for $S^x$, $S^y$, $S^z$ are related by unitary rotations that leave the Frobenius norm invariant, so their profiles coincide. Similarly $(S^z)^2 = \tfrac{1}{3}S(S+1)I + Q_{zz}$, where $Q_{zz}$ is the $zz$-component of the rank-2 quadrupole tensor. The scalar part acts trivially on the edge subspace, so the entire distinguishability
comes from the rank-2 part.
This holds exactly: the scalar part is a multiple of the identity, and any
such operator gives $F_{\mathrm{edge}} = c\,I_K$ and hence
$\delta_{\mathrm{edge}}=0$ for every $N$ and every position
(Appendix~\ref{app:Channel decomposition and derivation}).
Since $(S^x)^2$ contains a different component of the same rank-2 tensor, its profile is identical.

(ii) \emph{Why the chain $(S^z)^2$ profile is flat at zero.} The quadrupole $Q_{zz}$ is rank-2, so by the selection rule its reach is controlled by the $\mathcal L=2$ sector. In the spin-1 chain the doubled virtual space is only four-dimensional, $D^2=4$, and decomposes as
$\tfrac{1}{2}\otimes\tfrac{1}{2} = (\mathcal L{=}0)\oplus(\mathcal L{=}1)$
(Eq.~\eqref{eq:chain_decomp}). No $\mathcal L\ge 2$ sector exists, so the
rank-2 component has no channel through which to propagate, and its
reduced matrix elements vanish at the $10^{-17}$ level
(Table~\ref{tab:H1}).
What survives is only the finite-size channel discussed above, which for the chain can be evaluated in closed form
(Appendix~\ref{app:chain-closed}),
\begin{equation}
\delta_{\mathrm{edge}}\bigl((S^z)^2\bigr)
  = \frac{2\sqrt{6}}{9}\;3^{-(N-1)},
\label{eq:chain_rank2_closed}
\end{equation}
independent of position.  This is $4.68\times10^{-10}$ at $N=20$ and drops
below double precision for $N \gtrsim 33$, so the flat line seen at $N=60$
in Fig.~\ref{fig:kl}(c) is the machine-precision floor rather than the signal
itself, giving an effective $\xi=\infty$.
A rank-2 probe is thus exactly blind to the spin-1 chain edge in the
thermodynamic limit.

(iii) \emph{Why the ladder $\sum_\text{leg}(S^z_\text{leg})^2$ and $\sum_\text{leg}(S^x_\text{leg})^2$ profile decay then plateaus.}
For the three-leg ladder, the $\mathcal L=2$ sector exists.
The probe is $P_{tb}$-even, so it couples to $(\mathcal L{=}2,+)$, with
$|\lambda_{(2,+)}/\lambda_0|=0.1579$ and $\xi(\mathcal L{=}2,+)=0.5418$,
while the reduced matrix elements in all other subsectors vanish at the
$10^{-16}$ level (Table~\ref{tab:H1}).  The rank-2 signal therefore
decays within a few rungs, and once it falls below the finite-size channel,
the profile levels off.
We extract $\xi$ by fitting the full profile to $\mathrm A\,e^{-\mathrm{dist}/\xi}+\mathrm C$,
which cleanly separates the exponential signal from the plateau.

The comparison between (ii) and (iii) is exactly the point at which the three-leg ladder becomes indispensable.
The spin-1 chain has a virtual space of a single spin-$\tfrac{1}{2}$, whose doubled space reaches only up to $\mathcal L=1$.
It therefore cannot support any rank-$\ge 2$ probe, and the selection rule is invisible beyond rank 1.
The three-leg ladder has a virtual space $(\tfrac{1}{2})^{\otimes 3}$, whose doubled space extends up to $\mathcal L=3$.
This is the minimal AKLT geometry in the nontrivial Haldane phase in
which a rank-2 probe has a genuine propagation channel, and the minimal one
of any kind in which the hierarchy
$\xi(\mathcal L{=}1)>\xi(\mathcal L{=}2)>\xi(\mathcal L{=}3)$ is observable.

The leg-exchange symmetry $P_{tb}$ of Sec.~\ref{sec:logical gates} adds a
further refinement that is again unique to the ladder. Since $P_{tb}$
commutes with $\mathbf{J}^2_{\mathrm{adj}}$, each $\mathcal L$-sector splits into
$P_{tb}$-even ($+$) and $P_{tb}$-odd ($-$) sub-sectors, and a probe of
definite $P_{tb}$ parity couples only to the sub-sector of matching parity
(Appendix~\ref{app:Channel decomposition and derivation}).
The rung operators are $P_{tb}$-even.
Table~\ref{tab:H1} confirms that $S^z_{\mathrm{rung}}$ has a nonvanishing
reduced matrix element only in $(\mathcal L{=}1,+)$ and $\sum_\text{leg}(S^z_\text{leg})^2$ only in $(\mathcal L{=}2,+)$, the remaining five
sub-sectors sitting at $10^{-16}$.
The single-leg operator
$S^z_{\mathrm{top}}$ is not a $P_{tb}$ eigenoperator. It splits into a
$P_{tb}$-even part $\tfrac{1}{2}(S^z_{\mathrm{top}} + S^z_{\mathrm{bot}})$
and a $P_{tb}$-odd part
$\tfrac{1}{2}(S^z_{\mathrm{top}} - S^z_{\mathrm{bot}})$, and is the only
probe considered here that couples to two sub-sectors, with reduced matrix
elements $0.6734$ in $(\mathcal L{=}1,+)$ and $0.4166$ in
$(\mathcal L{=}1,-)$.
Since $|\lambda_{(1,+)}/\lambda_0| = 0.4801 >
|\lambda_{(1,-)}/\lambda_0| = 0.2823$ (Appendix~\ref{app:sectors},
Table~\ref{tab:sector_full}), the even part governs the asymptotic reach,
and both $S^z_{\mathrm{top}}$ and $S^z_{\mathrm{rung}}$ decay with
$\xi(\mathcal L{=}1,+) = 1.3630$.
The odd part, with
$\xi(\mathcal L{=}1,-) = 0.7907$, contributes only at short distance.

The small residual differences between $\xi_{\mathrm{fit}}$ and $\xi_{\mathrm{theory}}$ in Table~\ref{tab:sectors} come from the fitting window.
At short distances, the sub-leading eigenvalues of the same sub-sector contribute a faster transient, and near the chain center the profile saturates. 
Restricting the fit to the interior can bring every probe to the sector prediction.

Therefore, two symmetry labels determine what a local probe can see.
The SO(3) tensor rank $\ell$ fixes the sector $\mathcal L=\ell$, and, for the
ladder, the $P_{tb}$ parity fixes the sub-sector. The decay length then
follows from Eq.~(\ref{eq:xi_L}), and the fitted values of
Table~\ref{tab:fits} agree with this prediction for every operator
considered. Neither the strength of the probe nor its microscopic form
enters.

Read as an error model, $\delta_{\mathrm{edge}}\sim
e^{-\mathrm{dist}/\xi(\mathcal L=\ell)}$ states that bulk noise is
suppressed exponentially at a rate set by its symmetry label. Rank-2
noise is the extreme case.
In the chain, no $\mathcal L\ge2$ channel
exists, so only the finite-size correction of Eq.~(\ref{eq:chain_rank2_closed}) survives and the probe is blind in the
thermodynamic limit.
In the ladder, the channel exists but is short, $\xi(\mathcal L{=}2)=0.5418$.
The protection is passive and only
approximate, consistent with the Eastin--Knill theorem~\cite{eastin2009restrictions}.
Moreover, a boundary probe ($\mathrm{dist}=0$) resolves the edge states at full strength, so no finite code distance is defined, and only the reach of bulk probes is suppressed.

\section{Conclusion and Discussion}
\label{sec:conclusion}
We have studied the three-leg AKLT ladder as an exactly solvable quasi-one-dimensional model, using the spin-1 chain and the two-leg ladder as reference cases.
The analysis proceeded along three lines. First, we confirmed that the odd-leg ladders lie in the nontrivial
Haldane phase and the even-leg ladder lies in the trivial phase.
The conclusion follows from the $H^{2}\!\bigl(SO(3)\times\mathbb{Z}_2,\;U(1)\bigr)$ within the classification framework of~\cite{chen2011classification,ChenXieGuZheng-Cheng,Pollmann2012} and is
corroborated by independent string-order and entanglement-spectrum data, consistent with the stacking rule of one-dimensional SPT order~\cite{Pollmann2012}.
Second, we derived the logical gates that follow from the on-site symmetry through the MPS intertwining mechanism. The $SO(3)$ symmetry yields a continuous family of logical rotations.
In addition, the geometric symmetry that exchanges the legs gives rise to a logical gate that implements a multi-qubit permutation.
Finally, we investigated the accessibility of single-site operators to the encoding subspace by computing the distinguishability measure $\delta_{\mathrm{edge}}$.
As single-site operators are moved from the edge toward the bulk, we found that their probing capability exhibits an exponential decay, with the decay length determined by the $SO(3)$ tensor rank, and we explained this hierarchy by deriving the corresponding selection rules.

There are two results specific to the ladder geometry.
The first is a logical gate generated by the geometric symmetry that exchanges the top and bottom legs. Its action corresponds to a qubit permutation, which has no analogue in a single AKLT chain.
The second is the $SO(3)$ tensor-rank selection rule for local edge probes. The three-leg ladder is the simplest AKLT system in which this rule can be exhibited.
A rank-$\ell$ single-site operator couples to the edge subspace only through the $\mathcal L=\ell$ sector of the transfer matrix. Consequently, its probing range into the bulk is governed by the sector-dependent correlation length $\xi(\mathcal L=\ell)$.

We stress the limited scope of these results. The gate set is not
universal, as expected for on-site symmetries of a one-dimensional SPT
phase~\cite{ stephen2017computational, else2012symmetry}.
The protection we discussed is passive.
It suppresses the influence of bulk noise on the encoded information, but it involves no syndrome measurement and no
recovery operation.
A boundary operator already resolves the encoded states, so no finite code distance is defined.
The purpose of the model is to make these two central features, a deterministically implementable gate set and passive error suppression, which the AKLT system possesses, explicit and exactly computable.

In the present analysis, $\delta_{\mathrm{edge}}$ diagnoses a single local operator.
The complete Knill--Laflamme condition, however, concerns a set of error operators $\{E_a\}$~\cite{knill1997theory}.
Different physical noise processes correspond to different choices of the error set $\{E_a\}$, such as dephasing and depolarizing channels described by local Kraus operators, as well as correlated multi-site noise.
This object is accessible with the same transfer-matrix machinery used here, since a product of local errors corresponds to inserting several $T_O$ factors along the chain.
The selection rule should also generalize.
A product of a rank-$\ell_1$ operator and a
rank-$\ell_2$ operator couples to the encoded subspace through the Clebsch--Gordan composition of the two tensors, that is, through the sectors $|\ell_1-\ell_2|\le \mathcal L\le \ell_1+\ell_2$~\cite{edmonds1996angular}.
We predict that, in the infinite-chain limit, $\delta_{\mathrm{edge}}$ also exhibits an exponential decay with increasing separation between two operators.
This gives a testable prediction that is richer than the single-operator rule, and it does not require any input beyond the tensor already constructed here.

Apparently, the same tensor-network method applies to wider and to differently connected ladders.
For five-leg and seven-leg ladders, such calculations would test whether the entanglement-spectrum degeneracy pattern, the odd--even rule, and the sector hierarchy persist for higher $\mathcal L$.
We expect no new phase, so this is a check of generality rather than a source of new physics. 
A different variant is to close the ladder into a tube along the leg direction.
The geometric symmetry is then a cyclic group rather than a single reflection, and its virtual representation is a cyclic shift of the leg registers.
This would enlarge the discrete part of the logical gate set beyond the leg-exchange gate found here. We leave these for later work.

\appendix

\section{Projectors in the parent Hamiltonian}
\label{app:parentH}

Each two-body term in the parent Hamiltonian projects a pair of physical spins onto its forbidden highest-spin sector. For two interacting spins with spin operators $\mathbf{S}_1$ and $\mathbf{S}_2$, the total-spin operator is defined as $\mathcal{J}^2=(\mathbf{S}_1+\mathbf{S}_2)^2$.
Since $\mathcal{J}^2=\mathbf{S}_1^2+\mathbf{S}_2^2
+2\mathbf{S}_1\cdot\mathbf{S}_2$, the projectors can be written as polynomials of $\mathbf{S}_1\cdot\mathbf{S}_2$.

For the top and bottom legs, where two neighboring spins have $S_1=S_2=3/2$, the forbidden sector is $J=3$:
\begin{equation}
P_{J=3}\big|_{S_1=S_2=3/2} = \frac{\mathcal J_a^2(\mathcal J_a^2-2)(\mathcal J_a^2-6)}{720}.
\end{equation}
For the middle leg, where two neighboring spins have $S_1=S_2=2$, the forbidden sector is $J=4$:
\begin{equation}
P_{J=4}\big|_{S_1=S_2=2}= \frac{\mathcal J_b^2(\mathcal J_b^2-2)(\mathcal J_b^2-6)(\mathcal J_b^2-12)} {20\cdot 18\cdot 14\cdot 8}.
\end{equation}
For the rung coupling between spin-$3/2$ and spin-$2$ degrees of freedom, the forbidden sector is $J=7/2$:
\begin{equation}
P_{J=7/2}\big|_{S_1=3/2,\,S_2=2}= \frac{(\mathcal J_c^2-\tfrac34)(\mathcal J_c^2-\tfrac{15}{4}) (\mathcal J_c^2-\tfrac{35}{4})} {(\tfrac{63}{4}-\tfrac34)(\tfrac{63}{4}-\tfrac{15}{4}) (\tfrac{63}{4}-\tfrac{35}{4})}.
\end{equation}
The corresponding total-spin operators are
\begin{align*}
\mathcal J_a^2 = \frac{15}{2} + 2\mathbf{S}_1\cdot\mathbf{S}_2,
\qquad \mathcal J_b^2 = 12 + 2\mathbf{S}_1\cdot\mathbf{S}_2,
\qquad \mathcal J_c^2 = \frac{39}{4} + 2\mathbf{S}_1\cdot\mathbf{S}_2.
\end{align*}
Substitution of these expressions gives each projector explicitly as a polynomial in $\mathbf{S}_1\cdot\mathbf{S}_2$.

\section{SVD null-space extraction of virtual representations}
\label{sec:appD}

In this appendix we describe the numerical procedure used to extract the virtual representation matrices $V(g)$ from the MPS tensor.

\subsection{Linear constraint system}

The intertwining relation for a symmetry operation $g$ reads
\begin{equation}
  u_g:A^{s}\to A^{s}_g = \sum_{s'} [u_g]_{ss'} A^{s'} = \alpha_g \, V(g) A^{s} V(g)^{-1},\qquad \forall\, s = 1,\dots,d_{\text{phys}}, 
\end{equation}
where $\alpha_g = \pm 1$ is a phase factor. Multiplying both sides on the right by $V$ gives
\begin{equation}
A^{s}_g \, V = \alpha_g \, V \, A^s.
\end{equation}
This is a linear equation in the $D^2$ unknown entries of $V$.
To make the linear structure explicit, we vectorize. Using the standard identity $\operatorname{vec}(XYZ) = (Z^\top \otimes X) \operatorname{vec}(Y)$, where $Z^\top$ denotes the transpose of $Z$.
Then the left-hand side becomes $(I_D \otimes A^{s}_g) \operatorname{vec}(V)$ and the right-hand side becomes $\alpha_g \, ((A^s)^\top \otimes I_D) \operatorname{vec}(V)$. Rearranging, the constraint for each $s$ is
\begin{equation}
\bigl[ (A^s)^\top \otimes I_D - \tfrac{1}{\alpha_g} I_D \otimes A^{s}_g \bigr] \operatorname{vec}(V) = 0.
\end{equation}
Stacking these $D^2$-dimensional row blocks for all $s=1,\dots,d_{\text{phys}}$ yields the constraint matrix
\begin{equation}
\label{eq:constraint matrix}
\mathsf M = \begin{pmatrix}
(A^1)^\top \otimes I_D - \alpha_g^{-1} I_D \otimes A^{1}_g \\
(A^2)^\top \otimes I_D - \alpha_g^{-1} I_D \otimes A^{2}_g \\
\vdots \\
(A^{d_{\text{phys}}})^\top \otimes I_D - \alpha_g^{-1} I_D \otimes A^{d_{\text{phys}}}_g
\end{pmatrix} \in \mathbb{R}^{\, d_{\text{phys}} D^2 \times D^2}.
\end{equation}
The solution $\operatorname{vec}(V)$ lies in $\ker(\mathsf M)$, the null space of $\mathsf M$.

For the three-leg AKLT ladder, $D = 8$ and $d_{\text{phys}} = 80$, so $\mathsf M$ has $80 \times 64 = 5120$ rows and $64$ columns.

The null-space vector $\operatorname{vec}(V)$ is determined only up to an overall scalar. Thus, we fix this ambiguity by requiring $\|V\|_{\text{op}} = 1$, that is, the largest singular value of $V$ equals unity. 

\subsection{Verification}

After extraction, we verify the intertwining relation by computing the maximum residual
\begin{equation}
\label{eq:maximum residual}
\epsilon = \max_s \, \| A^{s}_g - \alpha_g \, V A^s V^{-1} \|_\mathrm F \;/\; \max_s \|A^s\|_\mathrm F,
\end{equation}
where $\|\cdot\|_\mathrm F$ denotes the Frobenius norm.
The results are summarized in Table~\ref{tab:verification}.

\begin{table}[ht]
\caption{Verification of the intertwining relation for each symmetry generator.}
\label{tab:verification}
\begin{ruledtabular}
\begin{tabular}{lccc}
Generator & Null-space dim. & $\alpha_g$ & $\epsilon$ \\
\hline
$g_z$    & $1$ & $+1$ & $4.83 \times 10^{-16}$ \\
$g_x$    & $1$ & $+1$ & $1.86 \times 10^{-15}$ \\
$P_{tb}$ & $1$ & $+1$ & $7.01 \times 10^{-16}$ \\
\end{tabular}
\end{ruledtabular}
\end{table}

All errors are at machine precision, confirming the correctness of the extraction.

\subsection{A posteriori identification with Pauli tensor products}
\label{sec:appD4}

To identify the extracted matrices with known operators, we compare each $V(g)$ with all $4^3 = 64$ three-qubit Pauli tensor products $\{\sigma_{\mu_1} \otimes \sigma_{\mu_2} \otimes \sigma_{\mu_3}\}$ (where $\sigma_0 = I$, $\sigma_1 = \sigma^x$, $\sigma_2 = i\sigma^y$, $\sigma_3 = \sigma^z$) by computing
\begin{equation}
\delta(\mu_1,\mu_2,\mu_3) = \min_{c = \pm 1} \| V(g) - c \; \sigma_{\mu_1} \otimes \sigma_{\mu_2} \otimes \sigma_{\mu_3} \|_\mathrm F,
\end{equation}
where $\|\cdot\|_\mathrm F$ denotes the Frobenius norm. The results for the rotation generators are summarized in
Table~\ref{tab:pauli_match}.

\begin{table}[ht]
\caption{Identification of the virtual representation matrices $V(g)$
for the rotation generators.}
\label{tab:pauli_match}
\begin{ruledtabular}
\begin{tabular}{lccc}
Generator & Best match & $c$ & $\delta$ \\
\hline
$g_z$ & $\sigma^z \otimes \sigma^z \otimes \sigma^z$             & $-1$ & $2.44 \times 10^{-16}$ \\
$g_x$ & $\sigma^x \otimes \sigma^x \otimes \sigma^x$             & $-1$ & $4.38 \times 10^{-16}$ \\
\end{tabular}
\end{ruledtabular}
\end{table}

For $P_{tb}$, the best Pauli tensor product match has deviation 0.866, reflecting the fact that $\operatorname{SWAP}_{13}$ is not a simple tensor product of Pauli matrices. Instead, it is identified directly from its action on the computational basis:
\begin{equation}
V(P_{tb}) \, |\alpha_1 \alpha_2 \alpha_3\rangle = |\alpha_3 \alpha_2 \alpha_1\rangle,
\end{equation}
which we verify holds exactly at machine precision for all eight basis states.

\section{Group cohomology of \texorpdfstring{$SO(3)\times\mathbb{Z}_2$}%
         {SO(3) x Z2}}
\label{app:cohomology}

We present the details of
\begin{equation}
H^{2}\bigl(SO(3)\times\mathbb{Z}_2,\,U(1)\bigr)=\mathbb{Z}_2,
\tag{\ref{eq:H2full}}
\end{equation}
used in Sec.~\ref{sec:virtual-symmetry}.
Throughout this appendix, group cohomology is understood in the Borel sense with trivial group action on the coefficient
module $U(1)=\mathbb{R}/\mathbb{Z}$.

For a direct product $G=G_1\times G_2$ of compact groups, and for a divisible coefficient module such as $U(1)$, the
K\"unneth formula for Borel group cohomology reads~\cite{ChenXieGuZheng-Cheng,brown2012cohomology}
\begin{equation}
  H^{n}(G_1\!\times\! G_2,\,U(1))
  =\bigoplus_{p+q=n} H^{p}(G_1,U(1))\otimes_{\mathbb{Z}} H^{q}(G_2,U(1)).
  \label{eq:kunneth_app}
\end{equation}
The Tor correction terms that would appear for non-divisible coefficients
vanish here because $U(1)$ is an injective $\mathbb{Z}$-module.
Specializing to $n=2$, $G_1=SO(3)$, $G_2=\mathbb{Z}_2$, and noting
$H^{0}(G,U(1))=U(1)$, which acts as the identity in the tensor-product
structure of cohomology classes:
\begin{equation}
  H^{2}(G,U(1))
  =
  H^{2}\bigl(SO(3),U(1)\bigr)
  \;\oplus\;
  \bigl[H^{1}\bigl(SO(3),U(1)\bigr)\otimes H^{1}(\mathbb{Z}_2,U(1))\bigr]
  \;\oplus\;
  H^{2}(\mathbb{Z}_2,U(1)).
  \label{eq:kunneth_expanded}
\end{equation}
We evaluate each factor below.

(i) $H^{2}(SO(3),U(1))=\mathbb{Z}_2$.
The group $SO(3)$ is connected, compact, and semisimple.
Since $SU(2)$ is the universal covering group of $SO(3)$, there exists a surjective Lie group homomorphism $\phi: SU(2) \to SO(3)$ with kernel $\ker \phi = \{\pm I\} \cong \mathbb{Z}_2$. Therefore, $\pi_1(SO(3)) \cong \mathbb{Z}_2$.
Hence $\pi_1(SO(3))=\mathbb{Z}_2$.
For a connected, compact Lie group $G$ with universal cover $\widetilde{G}$, every Borel projective $U(1)$-representation of $G$ lifts to a linear representation of $\widetilde{G}$, and equivalence classes of central extensions of $G$ by $U(1)$ are in bijection with $\mathrm{Hom}(\pi_1(G),U(1))$~\cite{ChenXieGuZheng-Cheng,moore1964extensions,brown2012cohomology}.
Therefore
\begin{equation}
  H^{2}\bigl(SO(3),U(1)\bigr)
  \cong \mathrm{Hom}\bigl(\pi_1(SO(3)),\,U(1)\bigr)
  = \mathrm{Hom}(\mathbb{Z}_2,\,U(1))
  = \mathbb{Z}_2.
  \label{eq:H2SO3_app}
\end{equation}
The generator of $\mathrm{Hom}(\mathbb{Z}_2,U(1))$ maps the nontrivial
element to $-1\in U(1)$, which corresponds to half-integer-spin representations.
Integer-spin representations correspond to the trivial homomorphism.

(ii) {$H^{1}(SO(3),U(1))=0$}. The first cohomology $H^{1}(G,U(1))=\mathrm{Hom}(G,U(1))$ is the group
of continuous one-dimensional representations.
Any continuous homomorphism $\rho\!:SO(3)\to U(1)$ must factor through
the abelianization $SO(3)/[SO(3),SO(3)]$.
Since $SO(3)$ is semisimple, its commutator subgroup is $SO(3)$ itself,
so the abelianization is trivial.
Consequently $\rho$ must be the trivial map, and
\begin{equation}
  H^{1}\bigl(SO(3),U(1)\bigr)=0.
  \label{eq:H1SO3_app}
\end{equation}
This ensures that the mixed term in Eq.~\eqref{eq:kunneth_expanded}
vanishes:
$0\otimes H^{1}(\mathbb{Z}_2,U(1))=0\otimes\mathbb{Z}_2=0$.

(iii)$H^{1}(\mathbb{Z}_2,U(1))=\mathbb{Z}_2$. 
For completeness we record
\begin{equation}
  H^{1}(\mathbb{Z}_2,U(1))=\mathrm{Hom}(\mathbb{Z}_2,U(1))=\mathbb{Z}_2,
\end{equation}
with the nontrivial homomorphism mapping the generator $a$ to $-1\in U(1)$.
This factor is irrelevant to the present classification because it is
annihilated by the vanishing of $H^{1}(SO(3),U(1))$.

(iv) $H^{2}(\mathbb{Z}_2,U(1))=0$.
Let $\mathbb{Z}_2=\{e,a\}$ with $a^2=e$.
A normalized 2-cocycle $\omega\!:\mathbb{Z}_2\times\mathbb{Z}_2\to U(1)$
must satisfy $\omega(e,g)=\omega(g,e)=1$ for all $g$; the only free
parameter is $\omega(a,a)\in U(1)$.
A 2-coboundary generated by $\mu\!:\mathbb{Z}_2\to U(1)$ with
$\mu(e)=1$ evaluates to
\begin{equation}
  (d\mu)(a,a)=\frac{\mu(a)\,\mu(a)}{\mu(a^2)}=\mu(a)^2.
\end{equation}
Since $U(1)$ is divisible, the squaring map $z\mapsto z^2$ is surjective:
for any $\omega(a,a)\in U(1)$ one can choose $\mu(a)$ with
$\mu(a)^2=\omega(a,a)$.
Hence every cocycle is a coboundary, giving
\begin{equation}
  H^{2}(\mathbb{Z}_2,U(1))=0.
  \label{eq:H2Z2_app}
\end{equation}

\section{Transfer-matrix evaluation of the string order parameter}
\label{sec:appB}

In this appendix we derive the transfer-matrix formula used in Sec.~\ref{sec:string-order} for the thermodynamic-limit string order parameter.

Consider a uniform infinite MPS built from a single tensor $A_{LR}^{s}$. For any one-unit-cell operator $C$, the associated transfer matrix $T_C$ is defined as in Sec.~\ref{sec:CG and transfer matrix}. In particular, we will need the ordinary transfer matrix $T$, the endpoint transfer matrix $T_{S^z}$, and the string transfer matrix $T_{e}$.

We start from the finite-distance string correlator
\begin{equation}
O_{\text{str}}(m) = \langle S_0^z \prod_{n=1}^{m-1} e^{i\pi S_n^z} S_m^z \rangle.
\end{equation}

In a finite periodic chain of length $N$, the expectation value of this operator is expressed as a product of transfer matrices. If the operator string occupies unit cells $0,\dots,m$, then one obtains
\begin{equation}
O_{\text{str}}^{(N)}(m) = \frac{\operatorname{Tr}\left( T^{N_L} T_{S^z} T_{e}^{\,m-1} T_{S^z} T^{N_R} \right)}{\operatorname{Tr}(T^N)},
\end{equation}
with $N_L + (m+1) + N_R = N$.

Let $\lambda_0$ be the dominant eigenvalue of $T$, and let $\langle l|$ and $|r\rangle$ be the corresponding left and right eigenvectors, normalized by
\begin{equation}
\langle l | r \rangle = 1.
\end{equation}
In the thermodynamic limit, $T^m$ is dominated by the projector onto this leading eigenspace,
\begin{equation}
T^m \sim \lambda_0^m |r\rangle\langle l|, \qquad m\to\infty.
\end{equation}
Substituting this form into the numerator and denominator, and then taking $N_L, N_R \to \infty$, yields
\begin{equation}
O_{\text{str}}(m) = \frac{\langle l| T_{S^z} T_{e}^{\,m-1} T_{S^z} |r\rangle}{\lambda_0^{\,m+1}}.
\end{equation}
It is convenient to factor out the dominant scale $\lambda_0$ from the string transfer matrix. This gives the numerically stable expression
\begin{equation}
O_{\text{str}}(m) = \frac{\langle l| \, T_{S^z} \left(T_{e}/\lambda_0\right)^{m-1} T_{S^z} |r\rangle}{\lambda_0^2},
\label{eq.D6}
\end{equation}
which is the formula used in the main text and in the numerical implementation.

To extract the asymptotic behavior and transient corrections, we perform a spectral decomposition of the normalized string transfer matrix. Define
\begin{equation}
\widetilde{T}_{e} \equiv T_{e}/\lambda_0.
\end{equation}
Assuming $\widetilde{T}_{e}$ is diagonalizable, let $\{\mu_i\}_{i=0}^{D^2-1}$ denote its eigenvalues, ordered by decreasing modulus $|\mu_0| \geq |\mu_1| \geq \cdots$, with corresponding right and left eigenvectors $|q_i\rangle$ and $\langle w_i|$ satisfying
\begin{equation}
\widetilde{T}_{e} |q_i\rangle = \mu_i |q_i\rangle, \qquad \langle w_i| \widetilde{T}_{e} = \mu_i \langle w_i|, \qquad \langle w_i | q_j \rangle = \delta_{ij}.
\end{equation}
The spectral resolution of the $(m-1)$-th power reads
\begin{equation}\label{eq:spectral-power}
\widetilde{T}_{e}^{\,m-1} = \sum_{i=0}^{D^2-1} \mu_i^{\,m-1} |q_i\rangle \langle w_i|.
\end{equation}
Substituting Eq.~\eqref{eq:spectral-power} into Eq.~(\ref{eq.D6}) yields
\begin{equation}\label{eq:ostr-spectral}
O_{\text{str}}(m) = \sum_{i=0}^{D^2-1} x_i \, \mu_i^{\,m-1},
\end{equation}
where the coefficients are defined by
\begin{equation}\label{eq:ci-def}
x_i = \frac{\langle l| \, T_{S^z} |q_i\rangle \, \langle w_i| \, T_{S^z} |r\rangle}{\lambda_0^2}.
\end{equation}

Two distinct cases arise depending on the dominant eigenvalue $\mu_0$ of $\widetilde{T}_{e}$. In the Haldane (SPT) phase, the dominant eigenvalue satisfies $\mu_0 = 1$, and consequently
\begin{equation}
O_{\text{str}}(\infty) = \lim_{m\to\infty} O_{\text{str}}(m) = x_0.
\end{equation}
All subleading eigenvalues obey $|\mu_i| < 1$ for $i \geq 1$, so that Eq.~\eqref{eq:ostr-spectral} separates into
\begin{equation}\label{eq:ostr-decomp}
\;O_{\text{str}}(m) = O_{\text{str}}(\infty) + \sum_{i \geq 1} x_i \, \mu_i^{\,m-1}.\;
\end{equation}

\section{Entanglement spectrum from MPS fixed points}
\label{sec:appC}

In this appendix we derive the entanglement-spectrum formula
Eq.~\eqref{eq:ES-formula} and establish the symmetry properties of the
fixed-point matrices used in Sec.~\ref{sec:entanglement}.

\subsection{Schmidt decomposition in the thermodynamic limit}
\label{sec:appC-derivation}

Consider an infinite uniform MPS cut into left and right half-infinite
parts at a single virtual bond of dimension~$D$.  The many-body state
can be written as
\begin{equation}\label{eq:C-Schmidt}
|\Psi\rangle = \sum_{\alpha=1}^{D} |L_\alpha\rangle \otimes |R_\alpha\rangle,
\end{equation}
where $|L_\alpha\rangle$ and $|R_\alpha\rangle$ are the half-chain states
associated with virtual boundary index~$\alpha$.  These states are
generally not orthonormal; their overlaps define two Gram matrices:
\begin{equation}\label{eq:C-Gram}
\langle L_\alpha | L_\beta \rangle = \bigl(G_L\bigr)_{(\alpha \beta)}, \qquad
\langle R_\alpha | R_\beta \rangle = \bigl(G_R\bigr)_{(\alpha \beta)}.
\end{equation}
Both $G_L$ and $G_R$ are Hermitian positive semidefinite by construction.

Within the transfer-matrix formalism, these Gram matrices coincide with
the dominant fixed points of the transfer matrix~$T$, reshaped as
$D\times D$ matrices.  To see this, note that the overlap of two right
half-chain states is obtained by iterating the transfer matrix along the
semi-infinite chain to the right of the cut.  In the thermodynamic limit,
only the contribution from the dominant eigenvalue~$\lambda_0$ survives:
\begin{equation}
T^k \;\sim\; \lambda_0^k \,|r\rangle\langle l|, \qquad k\to\infty,
\end{equation}
so that the right Gram matrix is given by the reshaped dominant right
eigenvector~$|r\rangle$.  The same argument applied to the left
half-chain identifies $G_L$ with the reshaped dominant left
eigenvector~$\langle l|$.  These are precisely the fixed points
introduced in Sec.~\ref{sec:string-order} for the string-order
computation.

The reduced density matrix of the left half-chain is
\begin{equation}
\rho_L = \operatorname{Tr}_R |\Psi\rangle\langle\Psi|
= \sum_{\alpha,\gamma} \langle R_\gamma | R_\alpha \rangle\,
|L_\alpha\rangle\langle L_\gamma|
= \sum_{\alpha,\gamma} \bigl(G_R\bigr)_{( \gamma \alpha)}\,
|L_\alpha\rangle\langle L_\gamma|.
\end{equation}
Consider an eigenvector $|v\rangle = \sum_\beta c_\beta\,|L_\beta\rangle$
of~$\rho_L$.  Acting with $\rho_L$:
\begin{equation}
\rho_L |v\rangle
= \sum_{\alpha,\gamma,\beta} \bigl(G_R\bigr)_{( \gamma\alpha)}\,\bigl(G_L\bigr)_{( \gamma\beta)}\,c_\beta\,
|L_\alpha\rangle.
\end{equation}
The eigenvalue condition $\rho_L|v\rangle = \lambda|v\rangle$ therefore
reduces to the $D$-dimensional matrix equation
\begin{equation}\label{eq:C-eigproblem}
G_R G_L\,c = \lambda\,c.
\end{equation}
The Schmidt weights are the eigenvalues of~$G_R G_L$.  Since $G_R G_L$ and $G_LG_R$
share the same nonzero spectrum, one may equivalently use
\begin{equation}\label{eq:C-spec}
\operatorname{spec}(\rho_L) = \operatorname{spec}(G_R G_L) = \operatorname{spec}(G_LG_R).
\end{equation}
This is Eq.~\eqref{eq:ES-formula} of the main text.

\subsection{Commutation of the edge state with the virtual symmetry}
\label{app:commutation}

We prove that the product $G_L G_R$ of the left and right transfer-matrix fixed points commutes with the virtual $\mathrm{SO}(3)$ representation $V(g) = u_{1/2}(g)^{\otimes M}$.

The transfer matrix is defined as (Eq.~\ref{eq:transfer_T})
\begin{equation}
T = \sum_{s=0}^{d_{\mathrm{phys}}-1} A^{s}\otimes \overline{A^{s}},
\end{equation}
acting on the doubled virtual space of dimension $D^2$.
The left and right fixed points $G_L$ and $G_R$ are positive $D\times D$ matrices satisfying
\begin{equation}\label{eq:right-fp}
  \sum_s A^s G_R\, (A^s)^\dagger = \lambda_0\, G_R,
\end{equation}
\begin{equation}\label{eq:left-fp}
  \sum_s (A^s)^\dagger G_L\, A^s = \lambda_0\, G_L,
\end{equation}
where $\lambda_0$ is the leading eigenvalue of $T$~\cite{perez2006matrix,schollwock2011density}.
The entanglement spectrum of a half-infinite bipartition is given by the eigenvalues of $G_L G_R$, normalized such that $\mathrm{tr}(G_L G_R)=1$.

The continuous $\mathrm{SO}(3)$ symmetry implies the following condition on the MPS tensor (Eq.~\ref{eq:intertwining}):
\begin{equation}
\label{eq:sym-appendix}
u_g:A^{s}\to A^{s}_g = \sum_{s'} [u_g]_{ss'} A^{s'} = \alpha_g \, V(g) A^{s} V(g)^{-1}, 
\end{equation}
for all $g\in \mathrm{SO}(3)$, where $|\alpha_g|=1$.

We first show $[G_R,\,V(g)]=0$. Substituting Eq.~\eqref{eq:sym-appendix} into Eq.~\eqref{eq:right-fp}, the left-hand side becomes
\begin{equation}
\sum_s A^s G_R\,(A^s)^\dagger
  = \sum_s \bigl(\alpha_g^{-1}\,V(g)^{-1}\, A^{s}_g\, V(g)\bigr)\, G_R\, \bigl(\bar{\alpha}_g^{-1}\,V(g)^\dagger\, (A^{s}_g)^\dagger\, (V(g)^{-1})^\dagger\bigr).
\end{equation}
Since $u_g$ is unitary, summing over $s$ is equivalent to summing over $s'$, and one obtains
\begin{equation}
  \sum_s A^s G_R\,(A^s)^\dagger = V(g)^\dagger\Bigl(\sum_s A^s\, \bigl[V(g)\, G_R\, V(g)^\dagger\bigr]\, (A^s)^\dagger\Bigr) V(g).
\end{equation}
Comparing with Eq.~\eqref{eq:right-fp} and using the uniqueness of the dominant eigenvector for an injective MPS~\cite{perez2006matrix}, we conclude
\begin{equation}
  V(g)\, G_R\, V(g)^\dagger = G_R.
\label{eq.e14}
\end{equation}
An analogous argument applied to Eq.~\eqref{eq:left-fp} yields
\begin{equation}
  V(g)^\dagger\, G_L\, V(g) = G_L,
\label{eq.e15}
\end{equation}
which is equivalent to $V(g)\,G_L\,V(g)^\dagger = G_L$, i.e.\ $[G_L,\,V(g)]=0$.

Since both $G_L$ and $G_R$ commute with $V(g)$, their product satisfies
\begin{equation}
  V(g)\,(G_L G_R)\,V(g)^{-1} = \bigl(V(g)\,G_L\,V(g)^\dagger\bigr)\bigl(V(g)\,G_R\,V(g)^\dagger\bigr) = G_LG_R,
\end{equation}
which establishes $[G_LG_R,\,V(g)]=0$ for all $g\in\mathrm{SO}(3)$. 

\section{Verification of the $P_{tb}$ logical gate}
\label{app:ptb_verify}

In this appendix we collect the numerical verifications specific to the geometric-reflection logical gate $\mathcal O(P_{tb})$ discussed in Sec.~\ref{sec:logical gates}.
All results are obtained from the exact MPS tensor construction with bond dimension $D=8$, physical dimension $d_{\text{phys}}=80$, and transfer-matrix power $N=20$.

\subsection{Eigenvalue spectrum of $\mathrm{SWAP}_{13}$ and $O(P_{tb})$}
\label{app:40+24}

The operator $\mathrm{SWAP}_{13}$ acts on the three-qubit space $(\mathbb{C}^2)^{\otimes 3}$ by exchanging the first and third qubits. Its $+1$ eigenspace consists of states symmetric under this exchange:
\begin{equation}
\mathcal{V}_+ = \mathrm{span}\{|000\rangle,\;|010\rangle,\;|111\rangle,\;|101\rangle,\;\tfrac{1}{\sqrt{2}}(|001\rangle+|100\rangle),\;\tfrac{1}{\sqrt{2}}(|011\rangle+|110\rangle)\},
\end{equation}
of dimension~6, while the $-1$ eigenspace,
\begin{equation}
\mathcal{V}_- = \mathrm{span}\{\tfrac{1}{\sqrt{2}}(|001\rangle-|100\rangle),\;\tfrac{1}{\sqrt{2}}(|011\rangle-|110\rangle)\},
\end{equation}
has dimension~2. Since $\mathcal O(P_{tb}) = \mathrm{SWAP}_{13}\otimes\mathrm{SWAP}_{13}$, its eigenvalues on $\mathbb{C}^8\otimes\mathbb{C}^8$ are products of the eigenvalues on each factor. The multiplicity of $+1$ is $6\times 6 + 2\times 2 = 40$, and that of $-1$ is $6\times 2 + 2\times 6 = 24$. Numerical diagonalization of the $64\times 64$ gate matrix confirms these multiplicities exactly.

\subsection{Non-factorizability into single-qubit gates}
\label{app:single-qubit gates}

We verify that $\mathcal O(P_{tb})$ cannot be written as $U_1\otimes\cdots\otimes U_6$ by the following procedure. The $64\times 64$ matrix is reshaped into a rank-12 tensor with six input and six output indices, each of dimension~2. For each logical qubit~$q$, an effective single-qubit matrix is extracted by fixing all other indices in the $|0\rangle$ state:
\begin{equation}
[A_q]_{ab} = \bigl[\mathcal O(P_{tb})\bigr]_{0\cdots a_q\cdots 0,\;0\cdots b_q\cdots 0}.
\end{equation}
The best scalar approximation $\mathsf c\,(A_1\otimes\cdots\otimes A_6)$ is then optimized over $\mathsf c\in\mathbb{R}$, and $\|\cdot\|_\mathrm F$ denotes the Frobenius norm.
The resulting relative error,
\begin{equation}
\frac{\lVert \mathcal O(P_{tb})-c\,(A_1\otimes\cdots\otimes A_6)\rVert_\mathrm F}{\lVert \mathcal O(P_{tb})\rVert_\mathrm F} \approx 0.66,
\end{equation}
is of order unity, confirming that $\mathcal O(P_{tb})$ is a genuinely multi-qubit gate.

\section{SO(3)$\times\mathbb{Z}_2$ sector decomposition of the doubled
virtual space}
\label{app:sectors}
\setcounter{table}{0}

\subsection{Symmetry sectors of the transfer matrix}
\label{app:sector:casimir}

The MPS tensor $A^s$ satisfies the intertwining relation (Eq.~\ref{eq:intertwining})
\begin{equation}
  u_g:A^{s}\to A^{s}_g= \sum_{s'} [u_g]_{ss'} A^{s'} = \alpha_g \, V(g) A^{s} V(g)^{-1}, \qquad \alpha_g = \pm 1,
\end{equation}
where $u(g)$ is the physical representation and
$V(g) = u_{1/2}(g)^{\otimes 3}$ acts on the virtual space
$\mathcal{V} = (\mathbb{C}^2)^{\otimes 3}$.  Using the unitarity
of $u(g)$, one shows
\begin{equation}
  [V(g)\otimes V(g)]\; T\; [V(g)\otimes V(g)]^{-1} = T
  \qquad \forall\, g\in\mathrm{SO}(3).
  \label{eq:T_SO3_commute}
\end{equation}
Under a $2\pi$ rotation,
$V(2\pi) = (-1)^3\,\mathbf{1} = -\mathbf{1}$, so $V(g)$ is
projective.  However,
$V(2\pi)\otimes V(2\pi) = +\,\mathbf{1}$, ensuring that
$V(g)\otimes V(g)$ is a linear representation of SO(3).
Its irreducible components therefore carry integer angular momentum
only.

Let $J^a$ ($a=x,y,z$) denote the spin generators on
$\mathcal{V}\cong\mathbb{C}^D$. The infinitesimal form of
Eq.~\eqref{eq:T_SO3_commute} yields $[T,\, J^a_{\mathrm{adj}}] = 0$, where
the adjoint generators are
\begin{equation}
J^a_{\mathrm{adj}} = J^a \otimes I_D - I_D \otimes (J^a)^{\top},
\label{eq:Jadj_def}
\end{equation}
where $(J^a)^{\top}$ denotes the transpose of $J^a$. 
The adjoint Casimir
\begin{equation}
\mathbf{J}^2_{\mathrm{adj}} = \sum_{a=x,y,z} (J^a_{\mathrm{adj}})^2
\label{eq:J2adj}
\end{equation}
therefore also commutes with $T$,
\begin{equation}
[T,\; \mathbf{J}^2_{\mathrm{adj}}] = 0.
\label{eq:T_commutes_J2}
\end{equation}
Its eigenvalues are $\mathcal L(\mathcal L+1)$ with $\mathcal L=0,1,\ldots$, and each eigenspace carries a $(2\mathcal L+1)$-fold magnetic degeneracy with $M=-\mathcal L,\ldots,\mathcal L$.
By Schur's lemma, within the
$\mathcal L$-sector the transfer matrix acts only on the $n_\mathcal L$-dimensional
multiplicity space, with $n_0=5$, $n_1=9$, $n_2=5$, $n_3=1$, yielding at
most $n_\mathcal L$ distinct eigenvalues per sector.
Numerically, the commutator norm
$\|[T,\,\mathbf{J}^2_{\mathrm{adj}}]\|/\|T\|$ 
vanishes identically for the chain and is $8\times 10^{-16}$ for the ladder, confirming the exact block-diagonal structure.

\subsection{Spin-1 chain sectors}
\label{app:sector:chain}

The virtual space is spin-$\frac{1}{2}$ ($D=2$). The doubled space has
$D^2=4$ dimensions and decomposes as
\begin{equation}
\tfrac{1}{2} \otimes \tfrac{1}{2} = (\mathcal L{=}0) \;\oplus\; (\mathcal L{=}1).
\label{eq:chain_decomp}
\end{equation}
This gives a one-dimensional $\mathcal L=0$ sector and a three-dimensional $\mathcal L=1$ sector, and no $\mathcal L\ge 2$ sector. The transfer-matrix eigenvalues normalized
by $\lambda_0$ are shown in Table~\ref{tab:chain_sectors}.

\begin{table}[ht]
\caption{Transfer-matrix sector decomposition for the spin-1 AKLT chain.}
\label{tab:chain_sectors}
\begin{ruledtabular}
\begin{tabular}{ccccc}
Sector $L$ & dim & Multiplicity & $\lambda / \lambda_0$ & $\xi(\mathcal L)$ \\
\hline
0 & 1 & 1 & $+1.0000$ & --- \\
1 & 3 & 1 & $-0.3333$ & 0.9102 \\
\end{tabular}
\end{ruledtabular}
\end{table}

The single $\mathcal L=1$ eigenvalue gives $\xi_{\mathrm{bulk}} = 1/\ln 3
\approx 0.9102$, in agreement with Eq.~\eqref{eq:xi1}. The absence of an
$\mathcal L\ge 2$ sector is the reason a rank-2 probe cannot resolve the edge states
of the chain, as discussed in Sec.~\ref{sec:selection_rules}~(ii).

\subsection{Three-leg ladder sectors}
\label{app:sector:ladder}

The virtual space is $(\frac{1}{2})^{\otimes 3} \cong \frac{3}{2} \oplus
\frac{1}{2} \oplus \frac{1}{2}$ ($D=8$). The doubled space has $D^2=64$
dimensions and decomposes as
\begin{equation}
8 \otimes 8 = 5 \cdot(\mathcal L{=}0) \;\oplus\; 9 \cdot(\mathcal L{=}1) \;\oplus\;
5 \cdot(\mathcal L{=}2) \;\oplus\; 1 \cdot(\mathcal L{=}3),
\label{eq:ladder_decomp}
\end{equation}
with total dimension $5+27+25+7 = 64$, in agreement with the numerical
diagonalization. The leg-exchange symmetry $P_{tb}$ commutes with
$\mathbf{J}^2_{\mathrm{adj}}$ and refines each $\mathcal L$-sector into $P_{tb}$-even
($+$) and $P_{tb}$-odd ($-$) sub-sectors, listed in
Table~\ref{tab:sector_full}.

The bulk correlation length $\xi_{\mathrm{bulk}} = 1.3630$
(Eq.~\eqref{eq:xi3}) corresponds to the $(\mathcal L{=}1,+)$ sector. Higher sectors
have progressively shorter reach, $\xi(\mathcal L{=}2) = 0.5418$ and
$\xi(\mathcal L{=}3) = 0.2960$. Since a rank-$\ell$ probe couples $\mathcal L{=}0$ to $\mathcal L{=}\ell$, its profile decays with $\xi(\mathcal L{=}\ell)$. This hierarchy, $\xi(\mathcal L{=}1) > \xi(\mathcal L{=}2) > \xi(\mathcal L{=}3)$, is the quantitative basis for the distinguishability profiles of Sec.~\ref{sec:selection_rules}, and it is
present only because the ladder virtual space extends to $\mathcal L=3$.

\begin{table}[ht]
\caption{SO(3)$\times P_{tb}$ sector decomposition of the three-leg ladder
transfer matrix. ``Mult.'' is the number of independent eigenvalue copies,
each $(2\mathcal L+1)$-fold degenerate. The eigenvalues are normalized by
$\lambda_0$.}
\label{tab:sector_full}
\begin{ruledtabular}
\begin{tabular}{lcclc}
Sector & dim & Mult. & Leading $\lambda/\lambda_0$ & $\xi$ \\
\hline
$(\mathcal L{=}0,+)$ & 3 & 3 & $+1.0000,\; +0.1141,\; +0.0459$ & --- \\
$(\mathcal L{=}0,-)$ & 2 & 2 & $+0.0682,\; -0.0341$ & 0.3725 \\
$(\mathcal L{=}1,+)$ & 15 & 5 & $-0.4801,\; -0.1834,\; +0.1024, \ldots$ & 1.3630 \\
$(\mathcal L{=}1,-)$ & 12 & 4 & $-0.2823,\; +0.1506,\; +0.0541, \ldots$ & 0.7907 \\
$(\mathcal L{=}2,+)$ & 15 & 3 & $+0.1579,\; +0.0605,\; -0.0341$ & 0.5418 \\
$(\mathcal L{=}2,-)$ & 10 & 2 & $+0.0887,\; -0.0341$ & 0.4128 \\
$\mathcal (\mathcal L{=}3,+)$ & 7 & 1 & $-0.0341$ & 0.2960 \\
\hline
\multicolumn{5}{l}{Total: $3+2+15+12+15+10+7 = 64$.} \\
\multicolumn{5}{l}{Completeness:
$\|\sum_{\mathrm{sectors}} \Pi - I_{64}\|_\mathrm F = 1.03 \times 10^{-14}$.} \\
\end{tabular}
\end{ruledtabular}
\end{table}

\subsection{Channel decomposition and derivation of Eq.~(\ref{eq:dKL_sector})}
\label{app:Channel decomposition and derivation}

This appendix derives Eq.~(\ref{eq:dKL_sector}) from the spectrum of the transfer matrix.
Throughout, $\eta_\mathcal L\equiv|\lambda_\mathcal L/\lambda_0|$, so that
$\xi(\mathcal L)=-1/\ln\eta_\mathcal L$ by Eq.~(\ref{eq:xi_L}).
We write $\eta_1$ for the largest such ratio over all non-leading
eigenvalues.

\subsubsection{Spectral decomposition and assumptions}

We assume the following.

(A1) The MPS is injective and $T$ is diagonalizable, with $\lambda_0$
unique in modulus.

(A2) The eigenvectors are biorthogonal,
\begin{equation}
T=\sum_i\lambda_i\,|r_i\rangle\langle l_i|,
\qquad \langle l_i|r_j\rangle=\delta_{ij},
\label{eq:H-spec}
\end{equation}
with $|r_0\rangle=|r\rangle$ and $\langle l_0|=\langle l|$ the dominant
fixed points of Sec.~\ref{sec:bulk}.

(A3) Reshaping a doubled-space vector into a $D\times D$ matrix by
$[\hat v]_{ab}=v_{(ab)}$, the matrices $\hat r_0$ and $\hat l_0$ are
invertible.

\subsubsection{Exact channel decomposition}

Let $\hat F$ act on unit cell $k$, and let $\alpha=(L_\alpha,R_\alpha)$,
$\beta=(L_\beta,R_\beta)$ be the raw boundary labels of Eq.~(\ref{eq:F_raw}).
Inserting $\hat F$ into the contraction that leads to Eq.~(\ref{Eq:21}) gives
\begin{equation}
(F_{\rm raw})_{\beta\alpha}
=\big[\,T^{k}\,T_F\,T^{\,N-1-k}\,\big]
 _{(L_\alpha,L_\beta),\,(R_\beta,R_\alpha)} ,
\label{eq:H-Fraw}
\end{equation}
where the product of transfer matrices is an ordinary matrix product on
the doubled virtual space.
The row pair consists of the two left boundary indices and the column
pair of the two right ones.
Eigenvectors of $T^{k}$ therefore attach to the left boundary, and those
of $T^{N-1-k}$ to the right boundary.

Inserting Eq.~\eqref{eq:H-spec} twice,
\begin{equation}
F_{\rm raw}=\sum_{i,j}\lambda_i^{\,k}\,\lambda_j^{\,N-1-k}\,
[T_F]_{ij}\;\Omega_{ij},
\label{eq:H-chan}
\end{equation}
with reduced matrix elements and channel matrices
\begin{equation}
[T_F]_{ij}\equiv\langle l_i|T_F|r_j\rangle,
\qquad
\Omega_{ij}\equiv\mathcal R\big(|r_i\rangle\langle l_j|\big)
=\hat r_i^{\,\top}\otimes\hat l_j ,
\label{eq:H-Omega}
\end{equation}
where $\mathcal R$ denotes the reshuffling operation, and the first factor acting on the $L$ component of the boundary label and
the second on the $R$ component.
All sector pairs $(i,j)$ are present.
Setting $\hat F=\mathds {1}$ gives $T_F=T$ and reproduces the Gram matrix,
\begin{equation}
G_N=\sum_i\lambda_i^{\,N}\,\Omega_{ii}
=\lambda_0^{N}\,\Omega_{00}\big[1+O(\eta_1^{N})\big].
\label{eq:H-GN}
\end{equation}

\subsubsection{Normalization of the projection}

By assumption~(A3) the matrix $\Omega_{00}=\hat r_0^{\,\top}\otimes\hat l_0$ is
invertible, so Eq.~\eqref{eq:H-GN} fixes the scale of $\mathsf P$ in Eq.~(\ref{Eq:p}),
\begin{equation}
\mathsf P=\lambda_0^{-N/2}\big[\tilde{\mathsf{P}} +O(\eta_1^{N})\big],
\qquad
\tilde{\mathsf{P}}^{\top}\,\Omega_{00}\,\tilde{\mathsf{P}}=I_K ,
\end{equation}
with $\tilde{\mathsf{P}}$ invertible and independent of $N$.
Define the rescaled projected channels
\begin{equation}
\widehat\Omega_{ij}\equiv\lambda_0^{N}\,\mathsf P^{\top}\Omega_{ij}\mathsf P
=\tilde{\mathsf{P}}^{\top}\Omega_{ij}\tilde{\mathsf{P}}+O(\eta_1^{N}).
\end{equation}
These are $O(1)$, and nonzero for every $(i,j)$. $\widehat\Omega_{00}=I_K+O(\eta_1^{N})$.
Projecting Eq.~\eqref{eq:H-chan} with Eq.~(\ref{eq:Oedge_def}) and extracting
$\lambda_0^{N-1}$,
\begin{equation}
F_{\rm edge}=\frac1{\lambda_0}\sum_{i,j}
\Big(\frac{\lambda_i}{\lambda_0}\Big)^{\!k}
\Big(\frac{\lambda_j}{\lambda_0}\Big)^{\!N-1-k}
[T_F]_{ij}\,\widehat\Omega_{ij} .
\label{eq:H-Fedge}
\end{equation}

\subsubsection{The trace subtraction}

On $K\times K$ matrices with the Hilbert--Schmidt inner product
$\langle X,Y\rangle=\mathrm{Tr}(X^{\dagger}Y)$, let
\begin{equation}
\Pi_\perp X\equiv X-\frac{\mathrm{Tr}X}{K}\,I_K .
\end{equation}
Then $\Pi_\perp^2=\Pi_\perp=\Pi_\perp^{\dagger}$, so $\Pi_\perp$ is the
orthogonal projector onto traceless matrices, and Eq.~(\ref{eq:c_deltaedge}) reads
\begin{equation}
\delta_{\rm edge}(F)=\frac{1}{\sqrt K}\big\|\Pi_\perp F_{\rm edge}\big\|_\mathrm F
=\frac{1}{\sqrt K}\min_{z}\big\|F_{\rm edge}-z\,I_K\big\|_\mathrm F.
\label{eq:H-pi}
\end{equation}
$\Pi_\perp$ is linear.

An exact consequence follows.
If $\hat F=c\,\mathds 1$ then $T_F=cT$, hence $F_{\rm raw}=c\,\mathcal R(T^{N})=c\,G_N$, and Eq.~(\ref{eq:29}) gives $F_{\rm edge}=c\,I_K$ for every $N$ and every $k$.
By linearity of $\hat F\mapsto F_{\rm edge}$,
\begin{equation}
\delta_{\rm edge}\big(\hat F+c\,\mathds 1\big)
=\delta_{\rm edge}\big(\hat F\big).
\label{eq:H-quot}
\end{equation}
The identity component of a probe is strictly invisible.
Numerically, the identity probe gives
$\delta_{\rm edge}=2.4\times10^{-16}$ (chain) and $1.7\times10^{-15}$ (ladder), both at machine precision.
We therefore replace $\hat F$ by its traceless part $\hat F_0=\hat F-(\mathrm{Tr}\hat F/d_{\rm phys})\mathds  1$
at the outset, and decompose $\hat F_0$ into irreducible spherical tensors.

Physically, $\mathrm{Tr}\hat{F}$ corresponds to the scalar component with rank $\ell=0$.
It is proportional to the identity operator within the encoded subspace and therefore cannot distinguish different logical states.
As a result, it carries no encoded information and is exactly removed by $\Pi_{\perp}$.
The traceless part $\hat{F}_0$ can be decomposed into various $SO(3)$ irreducible tensor components $T_{\hat{F}}^{(\ell)}$ whose rank $\ell\geq1$.
Their physical effects are determined by the subsequent selection rules.

\subsubsection{Four classes of channels}

The position dependence in Eq.~\ref{eq:H-Fedge} sorts the channels
into four classes.

\emph{(a) $i=j=0$.}
$k$ independent, and $\widehat\Omega_{00}=I_K+O(\eta_1^{N})$.
Proportional to $I_K$, hence annihilated by $\Pi_\perp$.
For $\ell\ge1$ the rule of Sec.~\ref{app:H-rules} makes
$[T_{F}]_{00}=0$ as well, so this class is absent altogether.

\emph{(b) $i\neq0$, $j=0$.}
Weight $(\lambda_i/\lambda_0)^{k}$.
Decays with the distance to the left boundary, independent of $N$.

\emph{(c) $i=0$, $j\neq0$.}
Weight $(\lambda_j/\lambda_0)^{N-1-k}$, the mirror image of (b).

\emph{(d) $i\neq0$, $j\neq0$.}
With $a=|\lambda_i/\lambda_0|<1$, $b=|\lambda_j/\lambda_0|<1$,
\begin{equation}
a^{k}\,b^{\,N-1-k}\le\max(a,b)^{\,N-1}\le\eta_1^{\,N-1}
\qquad\text{for all }k .
\label{eq:H-d}
\end{equation}
The bound is uniform in position.
This class is exponentially small in the system size rather than in the
distance to the boundary.
If a single modulus occurs among all non-leading eigenvalues, as for the
chain, the class (d) contribution is exactly position
independent.

Collecting (b) and (c),
\begin{align}
\Pi_\perp F_{\rm edge}
=&\;\frac1{\lambda_0}\sum_{i\neq0}
\Big(\frac{\lambda_i}{\lambda_0}\Big)^{\!k}[T_F]_{i0}\,
\Pi_\perp\widehat\Omega_{i0}
\notag\\
&+\frac1{\lambda_0}\sum_{j\neq0}
\Big(\frac{\lambda_j}{\lambda_0}\Big)^{\!N-1-k}[T_F]_{0j}\,
\Pi_\perp\widehat\Omega_{0j}
+O(\eta_1^{\,N-1}).
\label{eq:H-coll}
\end{align}

\subsubsection{Selection rules}
\label{app:H-rules}

Using $\mathsf P\mathsf P^{\top}=G_N^{-1}$ and Eq.~\ref{eq:H-GN},
\begin{equation}
\mathrm{Tr}\,\widehat\Omega_{ij}
=\mathrm{Tr}\big(\hat r_0^{-1}\hat r_i\big)\,
 \mathrm{Tr}\big(\hat l_j\hat l_0^{-1}\big)+O(\eta_1^{N}).
\label{eq:H-trace}
\end{equation}
By Eqs.~(\ref{eq.e14}) and (\ref{eq.e15}), $\hat r_0$ and $\hat l_0$ commute with
$V(g)$, hence so do their inverses, and both lie in the $\mathcal L=0$ sector.
The doubled representation $W(g)$ acts on $D\times D$ matrices by conjugation and is unitary with respect to the Hilbert--Schmidt pairing, so eigenspaces with different $\mathcal L$ are mutually orthogonal.
Therefore
\begin{equation}
\mathrm{Tr}\big(\hat r_0^{-1}\hat r_i\big)=0\;\;(L_i\neq0),
\qquad
\mathrm{Tr}\big(\hat l_j\hat l_0^{-1}\big)=0\;\;(L_j\neq0),
\end{equation}
so that $\mathrm{Tr}\,\widehat\Omega_{i0}=O(\eta_1^{N})$ and
$\Pi_\perp\widehat\Omega_{i0}=\widehat\Omega_{i0}+O(\eta_1^{N})\neq0$.
The projector acts as the identity on classes (b) and (c); their
amplitudes are neither reduced nor cancelled.
The same statement gives $c(F)=O(\eta_1^{N-1})$ for a probe with no
identity component.
The identical argument applied to the pair
$(\widehat\Omega_{\ell0},\widehat\Omega_{0\ell})$ shows that these two
matrices are Hilbert--Schmidt orthogonal.

From Eq.~(\ref{eq:intertwining}), a direct computation gives
\begin{equation}
W(g)\,T_{\hat C}\,W(g)^{-1}=T_{\hat C_g},
\qquad
\hat C_g=\bar u_g^{\,\dagger}\,\hat C\,\bar u_g ,
\label{eq:H-cov}
\end{equation}
with $\bar u_g$ the complex conjugate of the physical representation.
Every irreducible representation of $SO(3)$ is self-conjugate, so
$\bar u_g$ is unitarily equivalent to $u_g$ and the map
$\hat C\mapsto\hat C_g$ preserves the spherical-tensor rank.
Hence $T_{\hat F^{(\ell)}}$ is an irreducible tensor operator of rank
$\ell$ with respect to $W(g)$, whose infinitesimal generators are the
$J^a_{\rm adj}$ of Eq.~(\ref{eq:Jadj_def}).
The selection rule then gives
\begin{equation}
[T_{\hat F^{(\ell)}}]_{ij}=0
\qquad\text{unless}\qquad
|\mathcal L_i-\ell|\le \mathcal L_j\le \mathcal L_i+\ell .
\label{eq:H-WE}
\end{equation}
For class (b), $\mathcal L_j=0$ forces $\mathcal L_i=\ell$; for class (c), $\mathcal L_i=0$ forces
$\mathcal L_j=\ell$.
A rank-$\ell$ probe therefore reaches the code space only through the
$\mathcal L=\ell$ sector.
Setting $\mathcal L_i=\mathcal L_j=0$ requires $\ell=0$, which is the statement made under class (a).

\subsubsection{\boldmath$P_{tb}$ parity}

For the ladder, $u_{P_{tb}}$ is a real permutation, so
Eq.~\eqref{eq:H-cov} gives
$W(P_{tb})T_{\hat F}W(P_{tb})^{-1}=\pi_F\,T_{\hat F}$ whenever
$u_{P_{tb}}^{\dagger}\hat F\,u_{P_{tb}}=\pi_F\hat F$.
Hence
\begin{equation}
[T_{\hat F}]_{ij}=0
\qquad\text{unless}\qquad
\pi_i\pi_j=\pi_F .
\label{eq:H-parity}
\end{equation}
Since $|r_0\rangle$ lies in $(\mathcal L{=}0,+)$, a $P_{tb}$-even probe couples
only to $(\mathcal L{=}\ell,+)$ and a $P_{tb}$-odd probe only to
$(\mathcal L{=}\ell,-)$.

Both rules are confirmed in Table~\ref{tab:H1}.
For every ladder probe at most two of the seven sub-sectors carry a nonvanishing reduced matrix element, and all others sit at the $10^{-16}$ level.
The single-leg operators are not $P_{tb}$ eigenoperators and are the only probes
coupling to two sub-sectors.

\begin{table}[t]
\caption{\label{tab:H1}%
Reduced matrix elements $[T_{F_0}]_{i0}$ grouped by sector
$(\mathcal L,\pi)$, for the traceless part of each probe. Only sectors
with a nonvanishing entry are listed; ``others'' is the largest value
over all remaining sectors for that probe. The largest single matrix element in the sector differs between the $S^{z}$ and $S^{x}$ versions of the same operator.
The quadrature sum over the sector, which
is what enters $\delta_{\rm edge}$, is the same for both, in accordance
with observation~(i) of Sec.~\ref{sec:kl_profiles}.}
\begin{ruledtabular}
\begin{tabular}{llcccc}
System & Probe & $\ell$ & sector & $|[T_{F_0}]_{i0}|$ & others \\
\hline
chain  & $S^{z}$                     & 1 & $\mathcal L{=}1$ & 0.6667 & $0$ \\
chain  & $(S^{z})^{2}$               & 2 & none    & ---    & $6.7\times10^{-17}$ \\
chain  & $(S^{x})^{2}$               & 2 & none    & ---    & $5.7\times10^{-17}$ \\
ladder & $S^{z}_{\rm rung}$          & 1 & $(1,+)$ & 0.7999 & $7.7\times10^{-16}$ \\
ladder & $S^{x}_{\rm rung}$          & 1 & $(1,+)$ & 0.7035 & $5.8\times10^{-16}$ \\
ladder & $\sum_\text{leg}(S^{z}_\text{leg})^{2}$ & 2 & $(2,+)$ & 0.7208 & $7.2\times10^{-16}$ \\
ladder & $\sum_\text{leg}(S^{x}_\text{leg})^{2}$ & 2 & $(2,+)$ & 0.7642 & $7.1\times10^{-16}$ \\
ladder & $S^{z}_{\rm top}$           & 1 & $(1,+)$ & 0.6734 & $7.7\times10^{-16}$ \\
       &                             &   & $(1,-)$ & 0.4166 & \\
ladder & $S^{x}_{\rm top}$           & 1 & $(1,+)$ & 0.6441 & $4.5\times10^{-16}$ \\
       &                             &   & $(1,-)$ & 0.4367 & \\
\end{tabular}
\end{ruledtabular}
\end{table}

\subsubsection{The decay law}

Let $\lambda_{\mathcal L=\ell}$ be the largest-modulus eigenvalue in the
$\mathcal L=\ell$ sector, restricted to the appropriate $P_{tb}$ subsector, whose
reduced matrix element does not vanish, and set
$\eta_\ell=|\lambda_{\mathcal L=\ell}/\lambda_0|$.
Retaining this eigenvalue in each sum of Eq.~\ref{eq:H-coll},
\begin{equation}
\Pi_\perp F_{\rm edge}
=\Big(\tfrac{\lambda_{\mathcal L=\ell}}{\lambda_0}\Big)^{\!k} B_{\rm L}
+\Big(\tfrac{\lambda_{\mathcal L=\ell}}{\lambda_0}\Big)^{\!N-1-k} B_{\rm R}
+\ \text{transients}\ +O(\eta_1^{\,N-1}),
\label{eq:H-two}
\end{equation}
where the transients come from the subleading eigenvalues of the same
sector, and the matrices
$B_{\rm L}\propto[T_F]_{\ell0}\widehat\Omega_{\ell0}$,
$B_{\rm R}\propto[T_F]_{0\ell}\widehat\Omega_{0\ell}$
Equation~\eqref{eq:H-chan} is exact and contains every sector pair.
are independent of $k$ and of $N$.
By Sec.~\ref{app:H-rules} they are Hilbert--Schmidt orthogonal, so the
relative sign of the two terms is immaterial and
\begin{equation}
\delta_{\rm edge}(k)=\frac{1}{\sqrt K}
\sqrt{\;\|B_{\rm L}\|_\mathrm F^{2}\,\eta_\ell^{\,2k}
+\|B_{\rm R}\|_\mathrm F^{2}\,\eta_\ell^{\,2(N-1-k)}\;}\;.
\label{eq:H-quad}
\end{equation}
This is why moduli, rather than signed eigenvalues, appear even though
$\lambda_{\mathcal L=1}$ is negative in both models.
Inversion symmetry of the AKLT state gives
$\|B_{\rm L}\|_\mathrm F=\|B_{\rm R}\|_\mathrm F$, which makes
Eq.~\eqref{eq:H-quad} symmetric under $k\to N-1-k$.
Writing $\mathrm A\equiv\|B_{\rm L}\|_\mathrm F/\sqrt K$.
This is the reason the natural variable is $\mathrm{dist}(k)$ of Eq.~(\ref{eq:dist_def}).
For $1\ll\mathrm{dist}\ll N$ the first term dominates and
\begin{equation}
\delta_{\rm edge}(\mathrm{dist})\simeq
\mathrm A\,\eta_\ell^{\,\mathrm{dist}}
=\mathrm A\,e^{-\mathrm{dist}/\xi(\mathcal L=\ell)} ,
\label{eq:H-final}
\end{equation}
which is Eq.~(\ref{eq:dKL_sector}).

Table~\ref{tab:H2} confirms Eq.~\eqref{eq:H-coll} directly, by evaluating each class separately at $N=20$.
The class (d) contribution is uniform in position, as required by Eq~(\ref{eq:H-coll}).

\begin{table}[t]
\caption{\label{tab:H2}%
Channel-resolved $\delta_{\rm edge}$ at $N=20$ for the rank-2 probe of
the three-leg ladder.
Class (a) is at the $10^{-22}$ level and is not listed.
Since $\delta_{\text{edge}}$ is a Frobenius norm, the two orthogonal channel contributions add in quadrature; the squared fractions therefore sum to unity and directly reflect the relative weight of each channel.}
\begin{ruledtabular}
\begin{tabular}{cccccc}
$\mathrm{dist}$ & full & (b)+(c) & (d)
& $\left(\tfrac{\rm (b)+(c)}{\rm full}\right)^2$ & $\left(\tfrac{\rm (d)}{\rm full}\right)^2$ \\
\hline
0 & $1.1026\times10^{0}$  & $1.1026\times10^{0}$  & $1.4263\times10^{-6}$ & 1.0000 & 0.0000 \\
1 & $1.5408\times10^{-1}$ & $1.5408\times10^{-1}$ & $1.3836\times10^{-6}$ & 1.0000 & 0.0000 \\
2 & $2.3587\times10^{-2}$ & $2.3587\times10^{-2}$ & $1.3829\times10^{-6}$ & 1.0000 & 0.0000 \\
3 & $3.6920\times10^{-3}$ & $3.6920\times10^{-3}$ & $1.3825\times10^{-6}$ & 1.0000 & 0.0000 \\
5 & $9.1721\times10^{-5}$ & $9.1710\times10^{-5}$ & $1.3823\times10^{-6}$ & 0.9998 & 0.0002 \\
6 & $1.4543\times10^{-5}$ & $1.4477\times10^{-5}$ & $1.3823\times10^{-6}$ & 0.9910 & 0.0090 \\
7 & $2.6713\times10^{-6}$ & $2.2858\times10^{-6}$ & $1.3823\times10^{-6}$ & 0.7322 & 0.2678 \\
8 & $1.4287\times10^{-6}$ & $3.6094\times10^{-7}$ & $1.3823\times10^{-6}$ & 0.0638 & 0.9362 \\
9 & $1.3835\times10^{-6}$ & $5.7703\times10^{-8}$ & $1.3823\times10^{-6}$ & 0.0017 & 0.9983 \\
\end{tabular}
\end{ruledtabular}
\end{table}

\subsubsection{Closed form for the rank-2 probe of the chain}
\label{app:chain-closed}

For the spin-1 chain the class (d) residue can be written down exactly.

All non-leading eigenvalues of $T$ lie in the single $\mathcal L=1$
sector and share the modulus $1/3$ (Table~\ref{tab:chain_sectors}).
Every class (d) weight in Eq.~\eqref{eq:H-Fedge} is therefore
$(-1/3)^{\,N-1}$.
The traceless part of $(S^z)^2$ is the quadrupole $Q_{zz}=(S^z)^2-\tfrac23\mathds 1$.
A direct computation gives $T_{Q}(\mathds 1)=0$, which is the vanishing
entry of Table~\ref{tab:H1}, so classes (b) and (c) are absent.
On the $\mathcal L=1$ sector $T_{Q}/\lambda_0$ acts as
$\mathrm{diag}(2/9,\,2/9,\,-4/9)$ in the Cartesian basis
$\{\sigma^x,\sigma^y,\sigma^z\}$.
Assembling Eq.~\eqref{eq:H-Fedge} and taking the norm of
Eq.~(\ref{eq:c_deltaedge}) with $K=4$,
\begin{equation}
\delta_{\rm edge}\bigl((S^z)^2\bigr)
=\frac{1}{\sqrt4}\;2\sqrt{\Bigl(\tfrac29\Bigr)^{2}
+\Bigl(\tfrac29\Bigr)^{2}+\Bigl(\tfrac49\Bigr)^{2}}\;3^{-(N-1)}
=\frac{2\sqrt6}{9}\,3^{-(N-1)},
\label{eq:H-closed}
\end{equation}
independent of position, which is Eq.~(\ref{eq:chain_rank2_closed}).

The remaining single-sided channels survive the trace subtraction, are
restricted to $\mathcal L=\ell$ by Eq.~\eqref{eq:H-WE} and to one $P_{tb}$
subsector by Eq.~\eqref{eq:H-parity}, and add in quadrature.
The result is Eq.~\eqref{eq:H-final}, that is Eq.~(\ref{eq:dKL_sector}).

The scalar channel (a) is proportional to $I_K$ and is removed by the
trace subtraction; for $\ell\ge1$ it is absent altogether.
The channels (d), in which both transfer matrices are non-leading, are
bounded by $\eta_1^{\,N-1}$ uniformly in position. They are a
finite-size effect and vanish in the thermodynamic limit.
The remaining single-sided channels (b) and (c) survive the trace
subtraction, are restricted to $\mathcal L=\ell$ by
Eq.~\eqref{eq:H-WE} and to one $P_{tb}$ sub-sector by
Eq.~\eqref{eq:H-parity}, and add in quadrature.
This gives Eq.~\eqref{eq:H-final}, that is Eq.~(\ref{eq:dKL_sector}).

\begin{acknowledgments}
Y-N.Li is supported by National Natural Science Foundation of China under Grant No. 12005295.
\end{acknowledgments}

\bibliography{references}

\end{document}